\documentclass[usegraphicx]{mn2e}
\newcommand\etal{{\it et~al.}~}

\newcommand\kms{{\,km\,s$^{-1}$}}

\usepackage{longtable}
\usepackage{deluxetable}
\begin{document}

\title
[NGC\,5044 Group Membership and Dynamics]
{The anatomy of the NGC\,5044 group -- I. Group membership and dynamics}

\author
[Mendel \etal]
{J. Trevor Mendel\thanks{tmendel@astro.swin.edu.au},
Robert N. Proctor,
Duncan A. Forbes, 
Sarah Brough \\
Centre for Astrophysics \& Supercomputing, Mail H39, Swinburne University,
  Hawthorn, VIC 3122, Australia}

\maketitle

\begin{abstract}

We use a combination of new AAOmega multi-object wide-field spectroscopic observations and
literature data to define 111 spectroscopically confirmed members of the massive NGC\,5044 group
with $M_B\leq-13.5$\,mag, providing a three-fold increase in group members over previous analyses of
this group.  We find the group to have a dynamical mass of 9.2$\times 10^{14}$\,M$_{\odot}$, placing
it on the border between rich groups and poor clusters.  However, comparison to the
$L_X$--$\sigma_{\nu}$ and $L_X$--Mass relations shows it more closely follows cluster scaling
relations.  Using a combination of crossing time, X-ray contours and line-of-sight velocity profile
we are able to preclude growth of the NGC\,5044 group via major sub-group mergers within the last
$\sim$1\,Gyr.  While the majority of dynamical indicators for the group suggest it is virialized, we
find evidence for a small, dynamically distinct sub-group at 1.4\,Mpc from the group centre,
suggesting that the NGC\,5044 group is the dominant structure in its local environment, and is
currently accreting smaller groups.

\end{abstract}

\begin{keywords}
galaxies: clusters: NGC\,5044 group - galaxies: distances and redshift - galaxies: fundamental
parameters - galaxies: formation - galaxies: evolution 
\end{keywords}

\section{Introduction}
\label{intro}

It is well known that our own Galaxy is not isolated, residing with M31 and a host of smaller
galaxies in the Local Group.  Large area surveys have confirmed that the Local Group is not
exceptional, but rather representative of an extremely common environment (e.g.  Geller \& Huchra
1983; Eke \etal 2004; Weinmann \etal 2006).  However, despite the apparent commonality of the group
environment, the astrophysical processes acting within groups and their effects on observed galaxy
populations are relatively poorly understood.  This is due to both the low galaxy numbers and galaxy
surface densities typically associated with groups that make them observationally expensive.

Consequently, it is observations of galaxy clusters that have largely driven our current
understanding of environmental effects on galaxy populations, namely that galaxies in clusters are
redder (e.g.  Butcher \& Oemler 1984) and morphologically earlier types (e.g. Dressler 1980)
relative to galaxies in the field.  The expected correlation of star formation rate with density,
based on the morphology-density and morphology-radius relations, has led observers to measure star
formation rates as a function of environment in large samples such as the 2dFGRS (e.g. Lewis \etal
2002) and SDSS (e.g. G\'omez \etal 2003).  These authors have found that star formation rates
decrease as one moves from the field to cluster environment.  Perhaps more interestingly, however,
is that the observed star formation truncation is associated with local density rather than physical
proximity to a cluster, and therefore the physical mechanisms responsible are acting primarily in
lower density, group-like environments.  This has been confirmed by studies attempting to link field
and cluster galaxies, which find observed cluster populations cannot be generated directly from the
accretion of field galaxies.  Galaxy groups, therefore, {\it must} be playing a significant role as
an intermediary stage in galaxy evolution (e.g. Kodama \& Smail 2001; Fujita 2003).

After initial violent relaxation where galaxies are dominated by a collective potential (Lynden-Bell
1967), the dynamics of group and cluster galaxies are predominantly affected on more local scales.
Of the several possible mechanisms driving galaxy evolution, those dependent on the size of the
potential well (e.g. harassment; Moore 1996) or density of the intra-cluster medium (e.g.
ram-pressure stripping; Gunn \& Gott 1972) will be most effective in clusters.  Conversely, the
low-velocity group environment favours mergers as the preferred method of relaxation due to the
growing efficiency of dynamical friction at low relative velocities.

It has been shown that mergers can lead to both morphology and luminosity segregation (Fusco-Femiano
\& Menci 1998; Yepes, Dom\'inguez-Tenreiro \& Del Pozo-Sanz 1991), and observations suggest that
galaxies are segregated in both groups (Mahdavi \etal 1999; Girardi \etal 2003) and clusters (Adami,
Biviano \& Mazure 1998; Biviano \etal 2002; Lares, Lambas \& S\'anchez 2004).  Since mergers will
also affect the relative number of galaxies at a given luminosity, one might expect the prevalence
of mergers in groups to influence the shape of the group luminosity function.  This is, in fact, a
possible explanation of the commonly observed  ``dip'', indicative of intermediate mass galaxies
merging to form more luminous ones (e.g. Trentham \& Tully 2002; Miles \etal 2004).  It is also
worth noting that recent observations of ram-pressure stripping in groups (e.g. Bureau \& Carignan
2002; Kantharia \etal 2005; Rasmussen, Ponman \& Mulchaey 2006) support the results of numerical
simulations which suggest that group-level ram-pressure stripping could also play some role in group
galaxy evolution (Hester 2006). 

Typical studies of poor groups may classify 10-20 galaxies as group members, the majority of which
have had their membership assigned using photometric or morphological criteria.  More robust studies
have included only galaxies for which recession velocity measurements are available (e.g. Zabludoff
\& Mulchaey 1998; Carlberg \etal 2001; Brough \etal 2006a), however this often leads to the need to
stack groups in order to measure their properties due to the low numbers of redshifts typically
available. The problem with stacking groups, however, is that while it is then possible to constrain
the generalised global properties of groups (e.g. mass distribution, velocity dispersion profile
etc.) the individual properties of any single group are washed out.

Here we aim to address this issue by establishing the properties of the NGC\,5044 group and its
galaxies independently.  By studying a single rich group we are able to examine evidence for
dynamical segregation, substructure and peculiarities in its dynamical properties that are washed
out when stacking multiple groups.  In this paper, we present new deep spectroscopic data that allow
us to spectroscopically confirm $\sim40$ new group members.  With the addition of these velocities,
we create a new list of 111 confirmed group members that then allows a comprehensive analysis of the
dynamical attributes of the NGC\,5044 group and its constituent galaxies out to nearly two virial
radii.  In future work we will use these data to examine the stellar populations of the group
galaxies in relation to their position in the group, HI gas properties, star formation rates and
dynamical properties.

This paper is organised as follows: in $\S$\ref{n5044} and $\S$\ref{data} we describe some general
properties of the NGC\,5044 group and the data set we have assembled.  $\S$\ref{fof} describes
the method we have used to select group members from our list of potential candidates.  The global
group properties are addressed in $\S$\ref{global}, and the properties related to individual galaxies
are summarised in $\S$\ref{galaxy}.  

Throughout this paper we assume $H_0$ = 70\kms\,Mpc$^{-1}$ where applicable and recession velocities
are quoted in terms of c$z$.  We adopt the distance modulus of Tonry \etal (2001), $(m-M)_0=32.31$
(28.99\,Mpc), measured using surface brightness fluctuations with corrections applied to adjust for
the improved Cepheid distance measurements of Jensen \etal (2003).  Magnitudes have been corrected
for galactic extinction using the dust maps of Schlegel, Finkbeiner \& Davis (1998).  Following
convention we use $R$ to denote 2D, projected radii and $r$ to indicate 3D, deprojected radii.

\section{The NGC\,5044 group}
\label{n5044}

We have selected the NGC\,5044 group as the focus of this work as it is our goal to describe the
properties of a {\it single} group, and its large galaxy population is sufficient to perform a
statistical analysis.  

Ferguson \& Sandage (1990, hereafter FS90; 1991, hereafter FS91) photometrically studied 7 nearby
groups and clusters, including NGC\,5044, constructing group luminosity functions and examining the
relative fractions of dwarf and giant galaxies.  In their work, FS91 show that the early-type dwarf
to giant ratio (EDGR) correlates nearly monotonically with richness.  The intermediate richness of
the NGC\,5044 group, means that its EDGR occupies the transition between groups and clusters.  They
also found evidence for a steep faint-end luminosity function in the NGC\,5044 group, however their
photometric selection is likely to contain significant background contamination.  

The NGC\,5044 group has also been studied as part of the Group Evolution Multi-wavelength Survey
(GEMS; Osmond \& Ponman 2004; Forbes \etal 2006) and so has been observed at both X-ray and HI
wavelengths (Osmond \& Ponman 2004; McKay \etal 2004; Kilborn \etal in preparation, hereafter K08).
The GEMS group sample was selected from the availability of ROSAT PSPC pointings.  Further X-ray
observations of the NGC\,5044 group have been carried out using both Chandra and XMM--Newton (Buote
\etal 2003; Tamura \etal 2003; Buote, Brighenti \& Mathews 2004).  These authors found the group to
have a cool-core component of $\sim$0.7\,keV within 10\,kpc (consistent with the kinetic temperature
of stars in NGC\,5044 itself), and a warmer 1.4\,keV outside of 40\,kpc (characteristic of the group
halo).  The large radial extent of X-ray emission ($r>$60\,kpc) and high $L_X/L_B$ ratio
($\mathrm{log}_{10}L_X/L_B=31.82\pm0.01$\,ergs s$^{-1}$L$_{\odot}^{-1}$; Osmond \& Ponman 2004) are
key indications that the hot gas is associated with a group sized potential rather than just
NGC\,5044 itself. 

\section{Sample selection and data}
\label{data}

\subsection{AAOmega observations}
\label{aao_data}

We used the photometrically determined NGC\,5044 group catalogue of FS90 as the target list for new
spectral observations.  The FS90 catalogue was constructed using photographic plates covering
$\sim$2.3\,deg$^2$ taken at the 2.5m duPont Telescope at the Las Campanas Observatory.  AAOmega's
3\,deg$^2$ field-of-view allowed us to observe all 162 objects in the FS90 NGC\,5044 group catalogue
simultaneously. 

FS90 used a B-band radial cut of 16$^{\prime\prime}$, estimated to be 27 mag arcsec$^{-2}$ at the
distance of NGC\,5044), to limit background contamination.  Group membership was then defined using
morphological criteria to differentiate between background galaxies and genuine group members.  The
FS90 sample for NGC\,5044 is complete to B$_T\sim$18, and complete at the 90\% level to B$_T\sim$19. 

Our spectroscopic observations of the NGC\,5044 group were obtained in 2006 March on the
25--27$^\mathrm{th}$ using the AAOmega multi-object spectrograph at the 3.9m {\it Anglo-Australian
Telescope} (AAT).  Spectra were taken using medium resolution 580V and 385R gratings, yielding
dispersions of 3.5\,\AA\,and 5.3\,\AA\,FWHM respectively and spectral coverage from 3700--8800\,\AA.    

Since the targets spanned a wide range in magnitude ($-19.3\leq$M$_B\leq-12.3$) observations were
divided into two separate plate configurations to limit the effects of scattered light from the
brightest sources.  Plate configurations used a total of 197 fibres, 162 allocated to targets and 35
placed on blank sky regions.  Bright (faint) target observations were carried out in a series of 20
(30) minute integrations, with a total of 3-4 such integrations per plate configuration, for a total
of 4 (17) hours of observation across the 3 observing nights.
 
Data were reduced using the {\sc 2dfdr} software supplied and maintained by the Anglo Australian
Observatory.  This software handles over-scan subtraction, identification of the fibre apertures,
fibre flat-fielding and throughput calibration, wavelength calibration, median sky subtraction and
co-addition of science frames.  Due to known sensitivity issues in the blue arm of AAOmega at the
time of observing ($\sim$50\% decrease from expected), many of the faintest objects were lost in the
co-addition process as the reduced throughput resulted in these fibres being read-noise dominated.

Recession velocities were measured using the Fourier cross-correlation routine, {\sc fxcor}, in {\sc
iraf}.  These recession velocity measurements were made for blue and red frames separately to verify
the spectral wavelength solutions and recession velocity measurements.  Cross-correlations were
performed using both absorption (standard stars) and emission (rest frame emission galaxy)
templates.  We preferentially use absorption line templates for cross-correlation, however in cases
where absorption lines provided a poor redshift solution, emission templates were used.  In this
way, redshifts where measured for 103 galaxies out of the total 162 galaxy sample of FS90.

\begin{figure}
\centering
\includegraphics[scale=0.35,angle=-90]{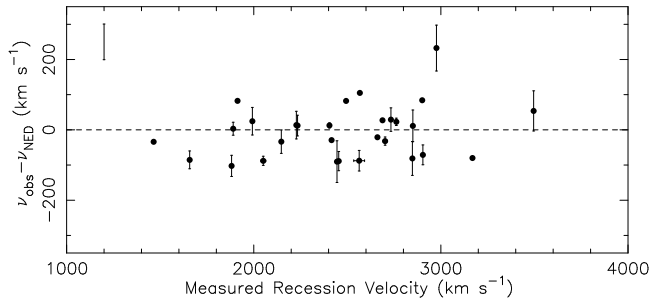}
\caption{Comparison of recession velocities measured in this work with available data from NED.
Only the 29 overlapping galaxies with measured recession velocities $\leq$4000\,\kms are shown.  The
error bar in the upper left side of the figure represents our adopted systematic uncertainty of
50.7\kms.}
\label{vel_comp}
\end{figure}

Fig. \ref{vel_comp} shows the relationship between newly measured redshifts and those available in
the NASA/IPAC Extragalactic Database (NED), showing the velocity difference between the two plotted
against our measured velocities.  Velocities are shown only for those galaxies for which we find
recession velocities $\leq$4000\kms (the range of interest for possible NGC\,5044 group members),
however background objects with higher velocities show similar good agreement with the literature.
We find no evidence for systematic variations with recession velocity out to a redshift of $\sim$0.7
(our highest redshift object), however based on the median absolute deviation between our velocity
measurements and those in NED we adopt a systematic uncertainty of 50.7\kms.

\subsection{6dFGS data}
\label{6df_data}

In order to include the most complete NGC\,5044 group member list available, we have used
supplementary data from the second data release of the 6dF Galaxy Survey (6dFGS DR2; Jones \etal
2005) to expand our AAOmega sample.  The 6dFGS is a large, 17046 deg$^2$ survey of southern sky
targets selected from the 2MASS Extended Source Catalogue (Jarrett \etal 2000) to include all
galaxies brighter than $K_{\mathrm{tot}}=12.75$\,mag.  To our list of possible group members we add all
6dFGS galaxies within 4 degrees of NGC\,5044 ($\sim$2\,Mpc in projected separation) and recession
velocities between 500 and 5000\kms, yielding 24 galaxies.  As of 6dFGS DR2 the region surrounding
NGC\,5044 has not been surveyed completely, and so 6dFGS galaxies included here are not uniformly
distributed. In particular, galaxies with declinations greater than $\sim -15.25^{\circ}$
($\sim$0.65\,Mpc north of the group centre) are not present in the 6dFGS DR2 sample (see Fig.
\ref{spatial} for a more detailed footprint of 6dFGS coverage in this region).
 
\subsection{GEMS HI data}
\label{kilborn}

The NGC\,5044 group has been surveyed for neutral hydrogen as part of the GEMS project using the
20-cm multi-beam receiver on the Parkes radio telescope by K08 (see also McKay \etal 2004).  They
surveyed a $5.5^{\circ}\times 5.5^{\circ}$ region centred roughly on NGC\,5044 and identified 23 HI
detections above a mass limit of $1.8\times10^8$\,M$_{\odot}$ (flux levels above 18 mJy
beam$^{-1}$).  Of these, 5 represent new galaxy detections which have been confirmed using high
resolution imaging from the Australian Telescope Compact Array (further details will be given in
K08), and so these new galaxies are included in our velocity sample.

\subsection{NED data}
\label{NED}

Given the incomplete spatial coverage available from the 6dFGS DR2 data we have supplemented these
data using sources with known recession velocities from the NED in the same position and velocity
range, providing 51 additional galaxies to our sample.  The heterogeneous nature of the NED data
means that an accurate estimate of incompleteness effects is difficult.  However, Brough \etal
(2006b) have shown NED to be complete in recession velocity and photometry to K$<$11 mag, with
statistically similar results obtained when fainter magnitude objects are included.

In total, the combined AAOmega, 6dFGS, GEMS HI and NED data provide 152 galaxies with redshifts
between 500 and 5000\kms in the region of NGC\,5044.

\subsection{Photometric data}
\label{phot_data}

In order to supply a cross-reference for the B-band magnitudes of FS90, as well as supplementary
B-band photometry for galaxies not in the FS90 galaxy list (i.e. those selected from 6dFGS, K08 and
NED) we have adopted additional B-band photometric data from Paturel \etal (2000; hereafter P00).
They obtain B$_T$ from Digitised Sky Survey (DSS) plates and provide total B-band magnitudes for 65 of
the 162 galaxies in the FS90 NGC\,5044 group sample, 21 out of 24 of our 6dFGS selected galaxies, 36
of the 51 NED galaxies, and 1 out of 5 from the K08 sample.

\begin{figure}
\centering
\includegraphics[scale=0.45,angle=-90]{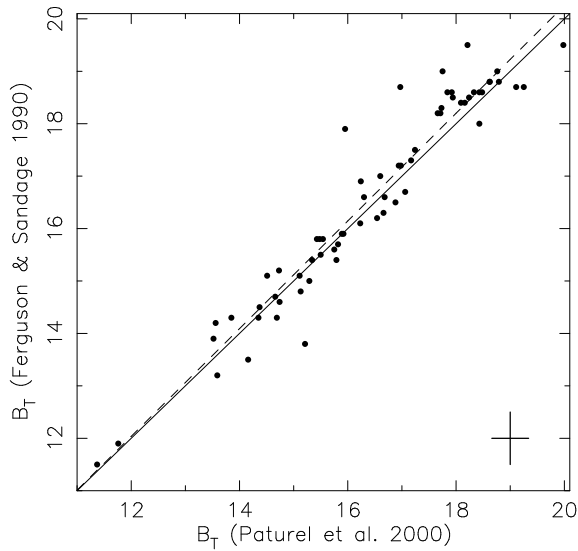}
\caption{Photometric comparison between B$_T$ as determined by FS90 and P00.  Errors for each point
are 0.5 and 0.34 mag as estimated by FS90 and P00 respectively, and are shown by the error bar in
the lower right corner.  The solid line represents the one-to-one line, while the dashed line
represents a least squares fit to the data.}
\label{b_phot_comp}
\end{figure}

In Fig. \ref{b_phot_comp} we show a comparison of FS90 and P00 total B-band magnitudes, showing the
good agreement of the two B$_T$ determinations.  There is no evidence for any systematic offsets
between the two samples.  The r.m.s. scatters about the one-to-one and best fit lines are 0.38 and
0.37\,mags respectively, and we therefore adopt the P00 magnitudes without additional correction when
FS90 measurements are unavailable.  

K-band magnitudes provide a reasonable proxy for stellar mass as they are relatively unaffected by
dust extinction and young, hot stars.  We have therefore compiled the available K-band photometry of
our sample.  The Two Micron All Sky Survey (2MASS, Skrutskie \etal 2006) has provided an excellent
resource of K-band photometry for a large number of galaxies, and supplies total K-band magnitudes
and isophotal radii (r$_{\mathrm K20}$) for 41 of the 162 galaxies in the FS90 target list and all
24 6dFGS selected galaxies.

\begin{figure}
\centering
\includegraphics[scale=0.45,angle=-90]{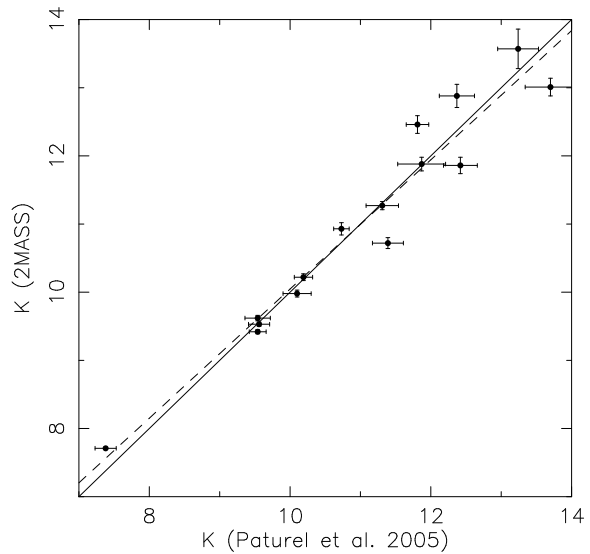}
\caption{Photometric comparison between K-band magnitudes from 2MASS and DENIS (as measured by P05).
Error bars represent 1-$\sigma$ deviations.  The solid line represents the one-to-one line, while
the dashed line represents a least squares fit to the data.}
\label{k_phot_comp}
\end{figure}

To provide greater K-band coverage, we adopt additional magnitudes from the Deep Near Infrared
Survey (DENIS; Paturel \etal 2005; hereafter P05).  In Fig. \ref{k_phot_comp} we show a comparison
of 2MASS and P05 K-band magnitudes.  The agreement between the two samples is good and there are no
suggestions of systematic offsets.  We preferentially use 2MASS K magnitudes, supplementing with P05
data where necessary.

\section{Group membership}
\label{fof}

Our AAOmega observations have allowed us to obtain new recession velocity measurements of a significant
number of galaxies for which we need to assess group membership.

Brough \etal (2006a; hereafter B06) used data from NED and 6dFGS to define the NGC\,5044 group using
a friends-of-friends algorithm (FoF; Huchra \& Geller 1982) which selects group members down to a
specified galaxy density above the critical density (in their case, $\delta$=150$\rho_0$).  Groups
derived using the FoF method are sensitive to the maximum allowed projected and velocity separation
(i.e. the linking length and velocity), the sample luminosity function and the magnitude limit of
the sample.  The inhomogeneous nature of our combined sample means that no single luminosity
function or magnitude limit will be adequate for selection of the group using an FoF algorithm, and
we therefore adopt a clipping method as it requires no {\it a priori} assumptions regarding our
data. 
 
Membership of the NGC\,5044 group is assigned using an iterated velocity dispersion clipping as
follows.  Galaxies are binned radially, using NGC\,5044 as the centre, and a moving mean velocity
and velocity dispersion, $\sigma$, are calculated.  Galaxies with velocities greater than 3$\sigma$
from the moving mean are removed from the group member list and the process is repeated until no
additional galaxies are removed.  

\begin{figure}
\centering
\includegraphics[scale=0.38,angle=-90]{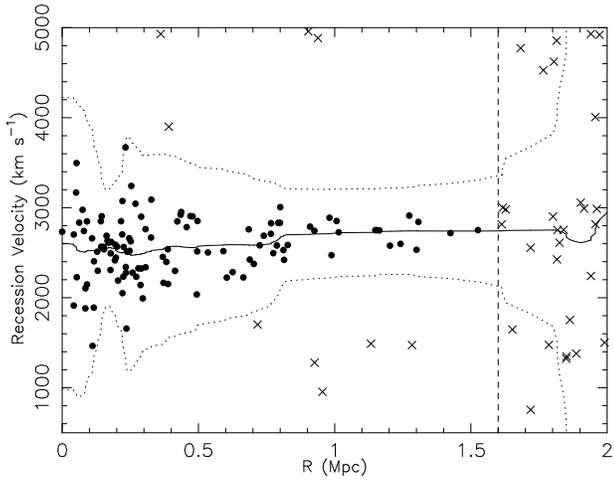}
\caption{Galaxy recession velocities against their projected distance from NGC\,5044.  Circles and
crosses represent galaxies included and excluded from the group respectively.  The vertical
dashed line shows a radial cut of 1.6\,Mpc, beyond which galaxies are no longer considered as
members.  Dotted lines represent the 3$\sigma$ velocity limit of the data, with the solid line
showing the moving mean velocity.}
\label{vel_radius}
\end{figure}

Velocity clipping is particularly well suited to this sample as the NGC\,5044 group is well
separated in velocity space from background contamination, illustrated in Fig. \ref{vel_radius}.  We
have imposed a cut at 1.6\,Mpc in projected radius, beyond which our moving velocity dispersion
begins to rise due to increasing velocity outliers.  It is likely that some of the galaxies beyond
this cut are also group members, however we are unable to reliably separate genuine group members
from contaminant galaxies, and so beyond this radius we no longer include galaxies as group members.
We note that the inclusion or exclusion of galaxies in this region has a negligible affect on the
dynamical parameters we derive for the group (i.e. mass, crossing time etc.).  

\begin{figure}
\centering
\includegraphics[scale=0.60,angle=-90]{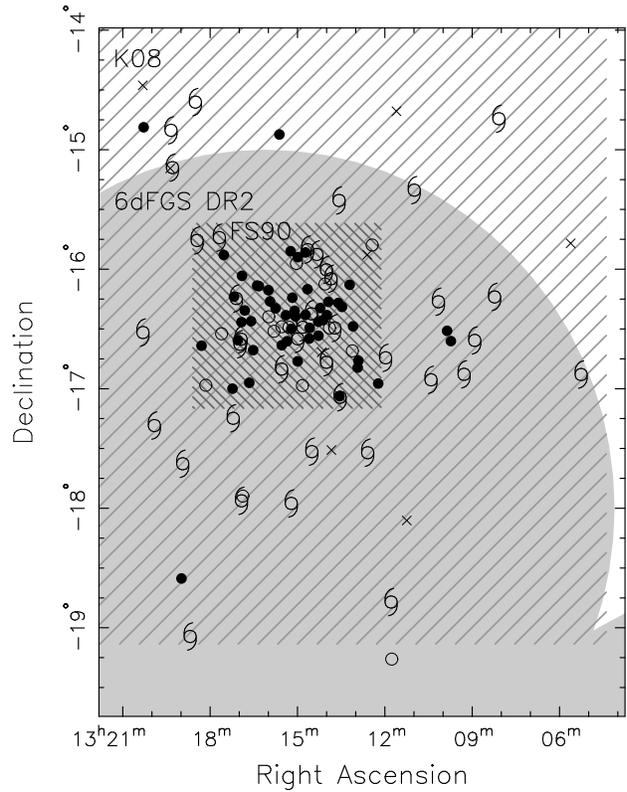}
\caption{Distribution of confirmed group members on the sky. Filled circles, open circles and
spirals represent elliptical, S0 and late type galaxies respectively.  Crosses mark group members
with unknown morphological types. The shaded and hashed regions indicate the areas covered by the
K08, FS90 and 6dFGS data.}   \label{spatial}
\end{figure}

The member list for the NGC\,5044 group is shown in Table \ref{group_gals} and includes 111 galaxies,
including 71 from FS90, 16 from 6dFGS, 19 from NED and 5 from K08. In Fig. \ref{spatial} we show the
distribution of our confirmed group member galaxies on the sky, with symbols representing different
morphological types.  

The group properties we find using velocity clipping rather than FoF are consistent within
uncertainties with those from B06.

\clearpage
\begin{center}
\begin{deluxetable}{lllccccl}
\tabletypesize{\footnotesize}
\tablecaption{Confirmed NGC~5044 Group Galaxies}
\tablewidth{0pt}
\tablehead{
\colhead{Galaxy Identifier} & \colhead{RA} & \colhead{DEC} & \colhead{c$z$} & \colhead{FS90} &\colhead{B$_T$}  & \colhead{K$_T$} & \colhead{Morphology} \\
\colhead{} & \colhead{(J2000)} & \colhead{(J2000)} & \colhead{(\kms)} &\colhead{Memb.} & \colhead{(mag)} & \colhead{(mag)} \\
\colhead{(1)} & \colhead{(2)} & \colhead{(3)} & \colhead{(4)} & \colhead{(5)} & \colhead{(6)} &\colhead{(7)} &\colhead{(8)}
}
\startdata
FS90 001		&	13 12 13.80	&	-16 57 22.0	&	2036.4  &	1	&	18.4	&\ldots		&dE,N?  		\\
FS90 003		&	13 12 26.35	&	-15 47 52.3     &	2904.0  &       3       &       14.3	&9.34$\pm$0.03	&Sb?  			\\
FS90 005		&	13 12 54.50	&	-16 45 57.0     &	2454.2  &       1       &       13.2 	&9.27$\pm$0.03	&E			\\
FS90 007		&	13 12 56.10	&	-16 49 22.0     &	2398.3  &       1       &       19.2	&\ldots	        &dE			\\
FS90 009		&	13 13 05.49	&	-16 28 41.7     &	1992.6  &       1       &       14.5 	&10.58$\pm$0.06	&S0			\\
FS90 011		&	13 13 06.80	&	-16 41 00.0     &	2669.1  &       1       &       18.9	&14.30$\pm$0.46	&dSd(on edge)	 	\\
FS90 015		&	13 13 12.48	&	-16 07 50.2     &	2761.8  &       1       &       14.6 	&10.72$\pm$0.08	&E			\\
FS90 017		&	13 13 28.42	&	-16 18 52.9     &	2661.0  &       2       &       15.6 	&\ldots	        &S0pec$^a$		\\
FS90 018		&	13 13 32.24	&	-17 04 43.4     &	2847.8  &       1       &       14.3	&9.98$\pm$0.05	&Sab			\\
FS90 019		&	13 13 33.40	&	-17 03 46.0     &	2298.2  &       1       &       17.7 	&12.74$\pm$0.62	&dE			\\
FS90 020		&	13 13 36.00	&	-16 16 58.0     &	2341.1  &       1       &       18.9	&\ldots	        &dE,N			\\
FS90 022		&	13 13 43.10	&	-16 29 58.0	&	2703.0	&  	1 	&	18.9	&\ldots		&dE/ImV			\\
FS90 024		&	13 13 50.96	&	-16 05 13.7     &	2521.0  &       3       &       17.2 	&\ldots	        &Sm$^a$			\\
FS90 027		&	13 13 54.15	&	-16 29 27.4     &	2444.7  &       1       &       14.2 	&9.81$\pm$0.05	&SB0$_3$		\\
FS90 029		&	13 13 56.21	&	-16 16 24.3     &	2415.9  &       1       &       15.5	&11.78$\pm$0.13	&d:E			\\
FS90 030		&	13 13 59.52	&	-16 23 03.8     &	2493.3  &       1       &       15.7	&\ldots	        &dE$^a$			\\
FS90 031		&	13 14 00.47	&	-16 00 44.2     &	2278.7  &       3       &       17.0	&\ldots	        &Im$^a$      	 	\\
FS90 032		&	13 14 03.22	&	-16 07 23.2     &	2850.3  &       1       &       14.0	&\ldots	        &S0			\\
FS90 034		&	13 14 07.40	&	-16 25 35.8     &	2688.2  &       1       &       16.6 	&\ldots	        &dE$^a$			\\
FS90 038		&	13 14 13.24	&	-16 19 29.0     &	2564.9  &       2       &       17.3 	&13.61$\pm$0.25	&E M32--type	 	\\
FS90 040		&	13 14 00.30	&	-16 47 10.0     &	3044.5  &       1       &       16.3 	&\ldots	        &SdIV			\\
FS90 041		&	13 14 16.70	&	-16 33 24.0     &	2635.2  &       2       &       18.8 	&\ldots	        &dE			\\
FS90 042		&	13 14 17.36	&	-16 26 19.5     &	2567.0  &       1       &       15.9 	&12.48$\pm$0.19	&dE/dS0$^a$	 	\\
FS90 045		&	13 14 20.00	&	-15 53 01.0     &	2141.0  &       3       &       18.9	&\ldots	        &S or Im		\\
FS90 049		&	13 14 31.23	&	-16 22 47.6     &	1465.2  &       1       &       15.4 	&\ldots	        &ImIII			\\
FS90 050		&	13 14 34.86	&	-16 29 28.9     &	2404.6  &       1       &       15.8 	&12.12$\pm$0.12	&dE pec,N/BCDring	\\
FS90 051		&	13 14 35.84	&	-16 34 50.4     &	2849.7  &       1       &       16.5 	&13.57$\pm$0.29	&dE,N			\\
FS90 053		&	13 14 39.22	&	-15 50 36.0     &	2324.3  &       1       &       16.2 	&13.29$\pm$0.21	&BCD?			\\
FS90 054		&	13 14 39.40	&	-16 10 04.0     &	2905.3  &       1       &       16.7	&\ldots	        &dE(Huge)		\\
FS90 057		&	13 14 43.54	&	-16 22 54.9     &	2103.9  &       1       &       18.8	&\ldots	        &dE			\\
FS90 058		&	13 14 43.61	&	-15 51 34.4     &	2326.4  &       1       &       18.8 	&\ldots	        &dE/Im			\\
FS90 063		&	13 14 49.82	&	-16 58 24.2     &	2338.9  &       1       &       15.2 	&11.73$\pm$0.12	&dS0			\\
FS90 064		&	13 14 49.23	&	-16 29 33.7     &	2147.4  &       1       &       13.9 	&9.62$\pm$0.04	&SBa			\\
FS90 068		&	13 14 59.37	&	-16 35 25.1     &	1890.2  &       1       &       13.0	&8.67$\pm$0.08	&Sab(s)			\\
FS90 069		&	13 14 59.80	&	-16 46 15.0     &	2567.6  &       3       &       18.7 	&\ldots		&dE		   	\\
FS90 070		&	13 14 59.40	&	-15 53 57.0	&	2626.6 	&	1	&	17.4	&\ldots		&dE			\\
FS90 072		&	13 15 02.13	&	-15 57 06.5     &	2234.2  &       1       &       13.9 	&9.53$\pm$0.03	&S0			\\
FS90 075		&	13 15 04.08	&	-16 23 40.2     &	1913.4  &       1       &       15.9 	&12.24$\pm$0.16	&dE$^a$			\\
FS90 076		&	13 15 05.90	&	-16 20 51.0     &	2702.0  &       1       &       16.1 	&12.46$\pm$0.13	&d:E or E M32-type	\\
FS90 078		&	13 15 10.54	&	-16 14 19.6     &	2741.3  &       1       &       15.0	&11.14$\pm$0.09	&E			\\
FS90 079		&	13 15 12.71	&	-16 29 57.1	&	2834.0	&	1	&	16.1	&12.88$\pm$0.17	&E			\\
FS90 081		&	13 15 14.30	&	-15 50 59.0     &	2232.9  &       2       &       19.1	&\ldots	        &dE			\\
FS90 082		&	13 15 17.57	&	-16 29 10.2     &	3495.8  &       1       &       14.7 	&10.93$\pm$0.09	&S0/a			\\
FS90 083		&	13 15 21.60	&	-16 36 08.0     &	2660.2  &       1       &       17.4	&\ldots	        &dE			\\
FS90 084		&	13 15 23.97	&	-16 23 07.9     &	2733.1  &       1       &       11.9 	&7.71$\pm$0.02	&E			\\
FS90 094		&	13 15 32.04	&	-16 28 51.1     &	3168.9  &       1       &       15.8 	&11.88$\pm$0.10	&d:E/S0			\\
FS90 095		&	13 15 32.40	&	-16 38 12.0     &	2510.7  &       1       &       18.8	&\ldots	        &dE,N?			\\
FS90 096		&	13 15 32.70	&	-16 50 35.0     &	3670.0  &       3       &       18.5 	&\ldots	        &(interacting)?	 	\\
FS90 100		&	13 15 45.13	&	-16 19 36.6     &	2227.6  &       1       &       14.3	&10.22$\pm$0.05	&E			\\
FS90 102		&	13 15 48.52	&	-16 31 08.0     &	1881.6  &       1       &       13.7	&9.42$\pm$0.03	&S0(cross)		\\
FS90 105		&	13 15 56.80	&	-16 16 10.0	&	2847.6	&	1	&	19.0	&\ldots		&dE,N			\\
FS90 107		&	13 15 59.30	&	-16 23 49.8     &	2976.5  &       1       &       13.8 	&9.43$\pm$0.15	&S0			\\
FS90 108		&	13 15 59.90	&	-16 10 32.0     &	2299.7  &       1       &       17.4 	&\ldots	        &dE,N			\\
FS90 116		&	13 16 19.70	&	-16 08 30.0	&	2612.9	&	3	&	19.0	&\ldots		&dE,N			\\
FS90 117		&	13 16 23.11	&	-16 08 11.4     &	2308.1  &       1       &       15.8 	&11.86$\pm$0.12	&d:E,N			\\
FS90 121		&	13 16 31.30	&	-16 40 40.0	&	2191.0	&	1	&	18.6	&\ldots		&dE/Im			\\
FS90 123		&	13 16 35.62	&	-16 26 05.7     &	2537.9  &       1       &       18.3	&\ldots	        &dE,N?			\\
FS90 127		&	13 16 39.40	&	-16 57 03.0     &	3089.3  &       1       &       18.6 	&\ldots	        &dE			\\
FS90 133		&	13 16 55.43	&	-16 26 32.2     &	2589.1  &       2       &       18.6   	&\ldots	        &dE,N			\\
FS90 134		&	13 16 56.23	&	-16 35 34.7     &	2051.0  &       1       &       15.1 	&\ldots	        &Sm(interacting)	\\
FS90 135		&	13 16 54.40	&	-16 03 18.0     &	3242.2  &       1       &       18.6 	&\ldots	        &dE,N			\\
FS90 137		&	13 16 58.49	&	-16 38 05.5     &	1657.7  &       1       &       11.5 	&\ldots	        &Sb(s)I-II(int.) 	\\
FS90 138		&	13 17 03.10	&	-16 35 39.0     &	2277.0  &       2       &       19.0 	&\ldots	        &dE,N			\\
FS90 141		&	13 17 06.13	&	-16 15 07.9     &	2563.3  &       1       &       13.5 	&\ldots	        &Sd(on edge)		\\
FS90 144		&	13 17 10.80	&	-16 13 46.7     &	2511.0  &       1       &       14.8 	&11.27$\pm$0.06	&E? + dE?		\\
FS90 146		&	13 17 14.00	&	-16 59 59.0     &	2538.1  &       3       &       19.1	&\ldots	        &S or dE,N		\\
FS90 151		&	13 17 32.00	&	-15 52 51.0     &	2165.0  &       2       &       19.0	&\ldots	        &dE			\\
FS90 153		&	13 17 36.38	&	-16 32 25.4     &	2900.2  &       1       &       15.4 	&11.73$\pm$0.11	&d:S0			\\
FS90 155		&	13 17 40.90	&	-15 44 35.0     &	2922.0  &       1       &       17.4	&\ldots	        &Sm                    	\\
FS90 158		&	13 18 09.04	&	-16 58 15.6     &	2784.2  &       3       &       16.9 	&13.15$\pm$0.16	&dS0 or S0		\\
FS90 161		&	13 18 17.80	&	-16 38 25.0	&	2153.0	&	1	&	18.9	&\ldots		&dE			\\
2MASX J13085477-1636106	&	13 08 54.79	&	-16 36 10.4     &	2585.8	&  \ldots       &       16.37   &12.97$\pm$0.15	&\ldots		\\
2MASX J13094408-1636077	&	13 09 44.07	&	-16 36 07.6     &	2582.0	&  \ldots       &       13.87	&9.75$\pm$0.03	&S0+		\\
2MASX J13091671-1653115	&	13 09 16.74	&	-16 53 11.8     &	2420.9	&  \ldots       &       16.40 	&12.83$\pm$0.19	&\ldots		\\
2MASX J13095347-1631018	&	13 09 51.70	&	-16 30 55.7     &	2376.0	&  \ldots       &       \ldots	&11.52$\pm$0.11	&S0--		\\
2MASX J13100952-1616458	&	13 10 09.52	&	-16 16 45.6     &	2226.0	&  \ldots       &       17.23	&12.34$\pm$0.13	&\ldots		\\
2MASX J13102493-1655578	&	13 10 24.96	&	-16 55 57.5     &	2423.9	&  \ldots       &       13.94 	&10.11$\pm$0.09	&\ldots		\\
2MASX J13114576-1915421	&	13 11 45.77	&	-19 15 42.3     &	2751.3	&  \ldots       &       \ldots	&9.44$\pm$0.04	&(R$^{\prime}$)SB(r)a 		\\
2MASX J13114703-1847312	&	13 11 47.04	&	-18 47 31.3     &	2532.7	&  \ldots       &       15.67	&12.93$\pm$0.12	&S		\\
2MASX J13115849-1644541	&	13 11 58.49	&	-16 44 54.1     &	2906.9	&  \ldots       &       15.29 	&11.96$\pm$0.09	&\ldots		\\
2MASX J13123543-1732326	&	13 12 35.43	&	-17 32 32.7     &	2760.0	&  \ldots       &       13.14	&8.85$\pm$0.03	&SAB(s)c		\\
2MASX J13133433-1525554	&	13 13 34.34	&	-15 25 55.2     &	2503.0	&  \ldots       &       13.71	&10.85$\pm$0.09	&SAB(s)dm		\\
2MASX J13143041-1732009	&	13 14 30.42	&	-17 32 00.9     &	2517.1	&  \ldots       &       15.23  	&11.54$\pm$0.09	&\ldots		\\
2MASX J13151278-1758006	&	13 15 12.80	&	-17 58 00.4     &	3006.4	&  \ldots       &       14.61	&12.32$\pm$0.15	&SAB(s)d		\\
2MASX J13153736-1452209	&	13 15 37.38	&	-14 52 21.1     &	2710.0	&  \ldots       &       \ldots	&11.24$\pm$0.08	&\ldots		\\
2MASX J13164875-1620397	&	13 16 48.75	&	-16 20 39.7     &	2619.1	&  \ldots       &	15.45 	&11.60$\pm$0.12	&\ldots	\\
2MASX J13165533-1756417	&	13 16 55.35	&	-17 56 42.0     &	2529.7	&  \ldots       &       15.26   &11.94$\pm$0.09	&S0--a		\\
2MASX J13171239-1715162	&	13 17 12.40	&	-17 15 16.1     &	2514.1	&  \ldots       &       14.24  	&10.24$\pm$0.04	&SAB(r)bc			\\
2MASX J13182685-1545599	&	13 18 26.85	&	-15 46 00.5     &	2853.0	&  \ldots       &       \ldots	&11.57$\pm$0.08	&Sb		\\
2MASX J13183034-1436319	&	13 18 30.35	&	-14 36 32.0     &	2889.0	&  \ldots       &       \ldots	&10.48$\pm$0.06	&SA(r)c pec?		\\
2MASX J13184125-1904476	&	13 18 41.26	&	-19 04 47.7     &	2718.3	&  \ldots       &       15.77	&12.06$\pm$0.10	&Sa		\\
2MASX J13185909-1835167	&	13 18 59.09	&	-18 35 16.7     &	2578.0	&  \ldots       &       14.14	&10.12$\pm$0.04	&SA0--		\\
2MASX J13191752-1509252	&	13 19 17.52	&	-15 09 25.3     &	2832.0	&  \ldots       &       \ldots	&10.85$\pm$0.06	&\ldots		\\
2MASX J13192062-1450402	&	13 19 20.65	&	-14 50 40.3     &	2744.0	&  \ldots       &       13.14	&9.16$\pm$0.03	&SB(s)c? sp		\\
2MASX J13192221-1509232	&	13 19 22.22	&	-15 09 23.5     &	2832.0	&  \ldots       &       \ldots	&13.16$\pm$0.11	&\ldots		\\
2MASX J13201698-1448455	&	13 20 17.01	&	-14 48 46.2     &	2853.0	&  \ldots       &       \ldots	&11.33$\pm$0.06	&\ldots		\\
2MASXi J1320185-163215	&	13 20 18.50	&	-16 32 12.0     &	2285.0	&  \ldots       &       14.44	&11.13$\pm$0.06	&\ldots		\\
6dF j1311150-180610 	&	13 11 15.03	&	-18 06 09.6     &	2727.0	&  \ldots       &       16.16	&13.80$\pm$0.23	&\ldots	\\
6dF j1313501-173048 	&	13 13 50.08	&	-17 30 47.8     &	2223.3	&  \ldots       &       16.84	&\ldots		&\ldots		\\
GEMS\_N5044\_5		&	13 05 37.27	&	-15 46 58.5     &	2912	&  \ldots       &       \ldots	&\ldots	        &\ldots 	\\
GEMS\_N5044\_14		&	13 08 05.02	&	-14 44 53.3     &	2599	&  \ldots       &       \ldots	&\ldots	        &\ldots 	\\
GEMS\_N5044\_7		&	13 10 59.34	&	-15 20 44.3     &	2828	&  \ldots       &       17.68	&13.62$\pm$0.23	&\ldots 		\\
GEMS\_N5044\_18		&	13 11 36.13	&	-14 40 40.1     &	2472	&  \ldots       &       \ldots	&\ldots	        &\ldots 	\\
GEMS\_N5044\_1		&	13 14 09.93	&	-16 41 41.6     &	3074	&  \ldots       &       \ldots	&\ldots	        &\ldots 	\\
PGC 045257		&	13 05 15.60	&	-16 53 19.0     &	2842	&  \ldots       &       15.05	&\ldots	        &SAB(s)m	\\
PGC 046242		&	13 16 52.70	&	-17 53 54.4     &	2580	&  \ldots       &       16.96	&\ldots	        &Dwarf	\\
PGC 046402		&	13 18 56.50	&	-17 38 06.0     &	2495	&  \ldots       &       16.31	&\ldots	        &SAB(rs)m	\\
PGC 046494		&	13 19 54.83	&	-17 18 55.8     &	2689	&  \ldots       &	15.54	&\ldots	        &SB(s)dm	\\
HIPASS J1312-15		&	13 12 36.20	&	-15 52 43.0     &	2953	&  \ldots       &       \ldots	&\ldots	        &\ldots	\\
HIPASS J1320-14		&	13 20 18.80	&	-14 27 42.0     &	2750	&  \ldots       &       \ldots	&\ldots	        &\ldots	\\
FGC1563			&	13 08 13.90	&	-16 14 21.0     &	2790	&  \ldots       &       16.4	&13.23$\pm$0.65	&Sdm		\\
\enddata
\tablecomments{Columns are as follows: (1) Galaxy Identifier (2) Right ascension (J2000), (3)
Declination (J2000), (4) Recession Velocity, (5) FS90 membership classification, where 1,2 and 3
represent definite, likely and possible members respectively, (6) B-band magnitude, (7) K-band
magnitude and error, (8) Morphological type.  For FS90 galaxies we adopt their morphological
classification, except where morphologies have been modified by Cellone \& Buzzoni (2005; marked with
$^a$).  For all other galaxies morphologies are adopted from NED.}
\label{group_gals}
\end{deluxetable}
\end{center}
\clearpage

\subsection{Comparison with FS90}
\label{redefine}

FS90 used morphological criteria to arrive at their original membership classification (e.g.
low-luminosity spiral galaxies are likely to be background objects).  Since we have obtained
recession velocities for 103 of their 162 NGC\,5044 group galaxies we are able to re-evaluate their
original group membership.  FS90 organised their observations into three different membership
classes designated 1 (definite members), 2 (likely members) and 3 (possible members).  

Our spectroscopic sample includes 60, 14 and 29 class 1, 2 and 3 galaxies respectively.  Confirmed
NGC\,5044 group members and their FS90 classifications are shown in Table \ref{group_gals}.
Non-members, their velocities and FS90 classifications are given in Table \ref{non_group}.
Of class 1 galaxies we find 55 of the 60 to be genuine group members, $\sim$92 percent, indicating
that class 1 galaxies in the FS90 catalogue are generally reliable group member galaxies.  The
fractions of genuine group member class 2 and 3 galaxies fall, as one might expect based on their
increasing uncertainty in group membership, and we confirm 7 class 2 and 9 class 3 galaxies as bona
fide group members (50 and $\sim$31 percent respectively).  We note that the NGC\,5044 group is the
most distant considered by FS90, and so it is perhaps not surprising that the classification of
galaxies is somewhat unreliable given the difficulty in morphologically classifying faint galaxies. 

\begin{figure}
\centering
\includegraphics[scale=0.36,angle=-90]{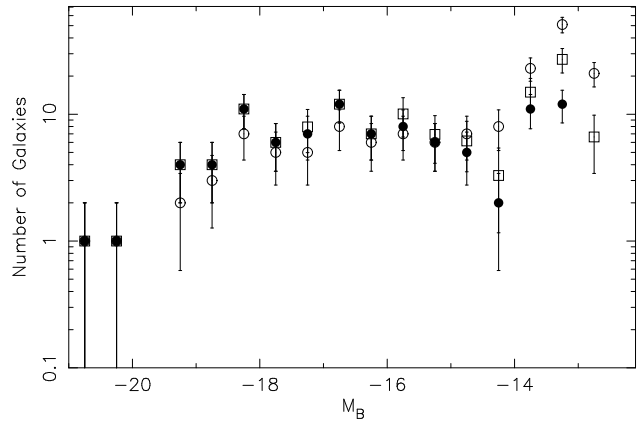}
\caption{B-band luminosity function for the NGC\,5044 group. Open circles represent the luminosity
function of the FS90 sample, while filled circles represent our spectroscopically confirmed group
member sample.  Open squares show the results of including additional galaxies from the FS90 sample
based on their likelihood of group membership (see $\S$\ref{redefine}).  Galaxies are binned at 0.5
magnitude intervals and Poisson errors are shown.  Note that points may overlap.}
\label{b_lumf}
\end{figure}

Fig. \ref{b_lumf} shows a comparison of the B-band luminosity functions (LF) between our confirmed
group sample and the complete FS90 sample.  The largest difference between the two comes at
faint magnitudes where our spectral observations become significantly incomplete and FS90 likely
includes a significant number of faint class 3 membership galaxies, of which our observations
suggest only one-third are actually group members.  

We adopt a Monte Carlo approach to correct the LFs shown in Fig. \ref{b_lumf} for our own
incompleteness and contamination present in the FS90 sample.  Confirmed group member galaxies are
always included in the analysis, however there are 59 dwarf galaxies in the FS90 sample with no
recession velocity information.  We therefore sample these remaining FS90 dwarf galaxies based on
the fractions of class 1, 2 and 3 included in our confirmed group sample (i.e.  92, 50 and 31
percent of each respective membership class are included in the Monte Carlo group each iteration).
This provides a more accurate estimate of the luminosity function and allows us to examine the
effect of contamination on the FS90 sample.

Fig. \ref{b_lumf} shows that our spectroscopic sample is in excellent agreement with the FS90 sample
for M$_B<-15$, where the FS90 sample is greater than 90 percent complete, and implies a flat
luminosity function.  In the region $-15<$M$_B<-14$ our spectroscopic sample begins to deviate from
the FS90 sample, however our Monte Carlo corrections show that the majority of this is due to the
inclusion of outliers in the FS90 sample, rather than incompleteness in our data.  Below M$_B>-14$
the LF shows tentative evidence for an upturn, however our sample is significantly incomplete in
this magnitude range so we abstain from drawing any further conclusions.
 
\begin{table}
\centering
\caption{Confirmed non-members and possible stars from the Ferguson \& Sandage (1990) NGC\,5044 Catalogue}
\begin{tabular}{rllrc}
\hline
FS90 & RA & DEC & c$z$ & FS90\\
(\#)   & (J2000) & (J2000) & (\kms) & Memb.\\
(1) & (2) & (3) & (4) & (5)\\
\hline
002  &  13 12 18.60  &  -16 47 40.0   &  69871.0   &    3  \\  
004  &  13 12 36.50  &  -16 13 52.0   &  4943.4    &    3  \\  
010  &  13 13 04.89  &  -16 15 46.9   &  12811.0   &    3   \\ 
021  &  13 13 39.60  &  -16 01 47.0   &  17082.3   &    3   \\ 
023  &  13 13 46.62  &  -16 20 07.9   &  103570.2  &    2   \\
028  &  13 13 55.90  &  -16 35 49.0   &  9054.7    &    1   \\ 
033  &  13 14 05.10  &  -15 55 42.0   &  13667.6   &    3   \\ 
037  &  13 14 12.30  &  -16 40 16.0   &  29078.4   &    3   \\ 
039  &  13 14 13.05  &  -16 11 14.9   &  28842.1   &    3   \\
046  &  13 14 25.59  &  -16 17 50.9   &  17122.0   &    1   \\ 
059  &  13 14 43.60  &  -15 44 48.0   &  42551.5   &    3   \\ 
088  &  13 15 26.80  &  -15 55 22.0   &  11048.0   &    3   \\ 
090  &  13 15 27.80  &  -16 11 41.0   &  28856.0   &    2   \\ 
099  &  13 15 42.70  &  -16 33 01.0   &  98838.7   &    3   \\ 
109  &  13 16 08.00  &  -16 59 49.0   &  5385.9    &    1   \\ 
110  &  13 16 06.79  &  -16 08 40.4   &  221397.9  &    3   \\ 
115  &  13 16 19.68  &  -16 40 13.7   &  5436.6    &    3   \\ 
124  &  13 16 35.80  &  -15 57 08.0   &  157159.7  &    2   \\
130  &  13 16 47.50  &  -16 28 04.0   &  17422.1   &    1   \\ 
136  &  13 16 56.68  &  -16 18 44.7   &  29325.4   &    3   \\ 
143  &  13 17 09.30  &  -16 51 18.0   &  84111.4   &    3   \\ 
152  &  13 17 34.30  &  -16 10 58.0   &  13925.5   &    2   \\ 
157  &  13 17 49.40  &  -16 53 39.0   &  19548.4   &    3  \\  
159  &  13 18 12.39  &  -15 41 25.8   &  11381.8   &    2   \\ 
160  &  13 18 11.50  &  -16 06 57.0   &  41112.6   &    3   \\ 
\multicolumn{5}{|c|}{\it possible stars}\\
008  &  13 12 57.30  &  -16 33 10.0   &  37.0 	& 3 \\  
014  &  13 13 12.20  &  -16 28 03.0   & $-$18.1 & 3 \\
026  &  13 13 54.70  &  -16 27 10.0   & $-$33.1 & 2 \\
036  &  13 14 12.30  &  -16 28 58.0   &  15.4  	& 2 \\
052  &  13 14 36.80  &  -16 07 56.0   & $-$71.6 & 3 \\
061  &  13 14 47.90  &  -16 17 51.0   & $-$49.7 & 3 \\
118  &  13 16 26.00  &  -17 03 19.0   &  87.2  	& 3 \\
\hline

\end{tabular}
\flushleft{Notes. Columns are as follows: (1) FS90 galaxy identifier, (2) Right ascension, (3)
Declination, (4) Measured recession velocity, (5) FS90 membership classification, where 1,2 and 3
represent definite, likely and possible members respectively.} 
\label{non_group}
\end{table}

\section{Global properties}
\label{global}

A vital step in establishing the role that groups play in the evolution of galaxies is establishing
their place between clusters and the field.  Are groups simply low-mass clusters, or do their member
galaxies exhibit unique signs of formation and evolution?  In order to assess this we first need to
establish the relation of the NGC\,5044 group to other groups and clusters.  

The dynamical properties of the NGC\,5044 group are summarised in Table \ref{lit_param}, which shows
the mean velocity, velocity dispersion, R$_{500}$ radius\footnote{The $R_{500}$ radius is defined as
the radius at which the projected density reaches 500 times the critical density.}, dynamical mass
and crossing time we derive using the formulae in Appendix \ref{Dynamical Formulae}. For comparison,
we also show similar measurements from the literature.

While we have more than tripled the number of group members included relative to previous studies,
our results for the mean velocity, velocity dispersion, R$_{500}$ radius, mass and crossing time are
all consistent within errors.  We find the group to be massive, nearly $10^{14}$\,M$_{\odot}$ in
total mass, which combined with the short crossing time leads us to the same conclusions of B06 that
the NGC\,5044 group is a classically massive, dynamically mature group.

\begin{table*}
\centering
\scriptsize
\caption{NGC\,5044 group dynamical properties compared with available literature values.  }
\begin{tabular}{lrccccc}
\hline
Source				&N	& $\bar{\nu}$	& $\sigma_{\nu}$& $r_{500}$	& $M_v$ 		& $t_c$	  	\\ 
				&	& (\kms)	& (\kms)	& (Mpc)		& (10$^{13}$M$_\odot$)	& ($H_0^{-1}$)	\\
(1)				&(2)	&(3)		&(4) 		&(5)		&(6)			&(7)		\\ 
\hline
This Work			&111	& 2577$\pm$33	& 331$\pm$26	&0.53$\pm$0.05	& 9.2$\pm$1.6 		& 0.03$\pm$0.004\\
B06				&32	& 2548$\pm$14	& 402$\pm$53	&0.55$\pm$0.07 	& 6.9$\pm$1.9		& 0.01$\pm$0.004 \\
Cellone \& Buzzoni (2005)	&26	& 2461$\pm$84	& 431		&\ldots		& \ldots		& \ldots \\
Osmond \& Ponman (2004)		&18	& 2518$\pm$100	& 426$\pm$74	&0.62$\pm$0.08	& \ldots		& \ldots \\
\hline
\end{tabular}
\flushleft Notes. Columns are as follows: (1) Source. (2) Number of group members included in the
calculated values. (3) Mean recession velocity and $\sigma$/$\sqrt{N}$ error. (4) Velocity
dispersion and 1$\sigma$ error (when available). (5) r$_{500}$ radius. (6) Crossing time.
\label{lit_param} 
\end{table*}

\subsection{Dynamical state}

Addressing the dynamical state of the NGC\,5044 group is key in our goal of establishing its
evolutionary history.  In particular, the brightest group galaxy (NGC\,5044) is known to have a
peculiar velocity of $\sim$150\kms with respect to the mean group velocity (Cellone \& Buzzoni 2005;
B06), which could be indicative of non-equilibrium due to recent mergers.  The peculiar velocity
found for NGC\,5044, scaled by the group's velocity dispersion, is 0.45$\sigma_\nu$, and within the
range typically found for relaxed groups and clusters (e.g. Oegerle \& Hill 2001; B06; Hwang \& Lee
2007).  Peculiar brightest group and cluster galaxy (BGG/BCG) velocities are therefore not in and of
themselves indicative of groups out of dynamical equilibrium. 

\begin{figure}
\centering
\includegraphics[scale=0.35,angle=-90]{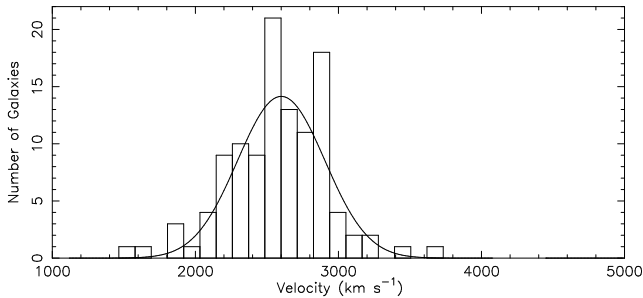}
\caption{Histogram of the velocities measured in our group sample.  The solid line
represents the best fit fourth order Gauss-Hermite series to the velocity data. }
\label{vel_profile}
\end{figure}

To further assess the dynamical state of the NGC\,5044 group we examine both the characteristics of
line-of-sight (LOS) velocities and dynamical measurements such as crossing time.  Fig.
\ref{vel_profile} shows the LOS velocity distribution for our group sample with the best fit
Gaussian overlayed.  Fitting a fourth-order Gauss-Hermite expansion to our LOS velocities (e.g.
Zabludoff, Franx \& Geller 1993) we find negligible values for the $h_3$ and $h_4$ coefficients,
measurements of the distribution skewness and kurtosis.  The normal distribution of LOS velocities
suggests that the group is relaxed and well mixed, as $h_3$ and $h_4$ are generally indicative of
substructure or strong orbital anisotropies in the group velocities (e.g. Merritt 1988; Zabludoff
\etal 1993). 

The crossing time of a group can also be used as a indication of a group's virialization, where
crossing times $> 0.09\,H_0^{-1}$ are usually indicative of groups whose galaxy orbits have not had
sufficient time to circularise in the group potential (Nolthenius \& White 1987).  The short
crossing time found for the NGC\,5044 group ($t_c=0.03\pm0.004\,H_0^{-1}$, $\sim$0.4\,Gyrs)
therefore suggests that it is most likely virialised and consistent with the low kurtosis we find
for the velocity distribution, discussed above.
 
The use of crossing time as an indicator of virialization can be somewhat questionable in groups,
and studies examining very small groups (N$_{gal}\sim5-10$) have found that using $t_c$ as the sole
indicator of virialization can lead to the inclusion of up to 20 percent spurious associations (see
e.g.  Diaferio \etal 1993; Niemi \etal 2007).  However this discrepancy is primarily a concern for
galaxies which are only questionably bound.  A comparison of the NGC\,5044 group to the virial
parameters discussed in Niemi \etal (2007) suggests that groups of order the NGC\,5044 group mass
are bound with greater than 99 percent probability and that the crossing time is a useful indicator
in assessing dynamical status.  

The availability of XMM X-ray observations for the NGC\,5044 group (T. Ponman, private
communication) means that we have an additional and independent measure for the group centre of mass.
Poole \etal (2006) have shown that an offset in the X-ray peak from the centroid of emission is an
excellent indicator of dynamical equilibrium in clusters as sub-cluster mergers can cause the X-ray
peak to ``slosh'' with respect to the centroid.  Using an XMM mosaic, we calculate the
peak--centroid displacement for the NGC\,5044 group, finding a nominal offset of
$\sim$0.003\,$R_{500}$, which is within the range of relaxed clusters in the Poole \etal (2006)
simulations.  We do note, however, that this test is sensitive only to offsets in the plane of the
sky, and so we are unable to rule out significant displacements of the X-ray peak along the
line of sight.

From the above indicators we therefore conclude that the NGC\,5044 group is in equilibrium and
virialized.
 
\subsection{Substructure} 
\label{sub_struct} 

Based on the paradigm of hierarchical structure formation, conglomerations of galaxies such as
clusters and large groups are formed via the successive infall of smaller structures.  Substructures
are found in anywhere from 30 to 70 percent of optically observed clusters (West 1988; Girardi \etal
1997; Burgett \etal 2004; Ramella \etal 2007), providing considerable evidence in support of this
merging scenario.  While the NGC\,5044 group is considerably smaller than rich clusters typically
analysed we still expect it to have evolved via similar hierarchical mechanisms, and so a search for
dynamical substructures is relevant.

Here we adopt the Dressler-Shectman $\Delta$ statistic (Dressler \& Shectman 1998) as a means for
measuring the three-dimensional (space-velocity) substructure present in our sample.  This
statistic, while originally used for measurement of substructure in clusters, has been shown to be
effective at identifying dynamical substructures even at smaller, group scales (Pinkney \etal 1996).

Calculation of the $\Delta$ statistic involves the summation of local velocity anisotropy, $\delta$,
for each galaxy in the group, defined as    

\begin{equation}
\label{delta}
\delta^2 =
\left(\frac{N_{\mathrm{nn}}+1}{\sigma^2}\right)[(\bar{\nu}_{\mathrm{local}}-\bar{\nu}_{\mathrm{group}})^2-(\sigma_{\mathrm{local}}-\sigma)^2]
\end{equation}

where $N_{\mathrm{nn}}$ is the number of nearest neighbours over which the local recession velocity
($\bar{\nu}_{\mathrm{local}}$) and velocity dispersion ($\sigma_{\mathrm{local}}$) are calculated.
Here we adopt $N_{\mathrm{nn}}=\sqrt{N}$, following Pinkney \etal (1996).  The test statistic
$\Delta$ is larger for increasing levels of substructure in the group.  

In order to provide a normalisation for this statistic, and therefore a means of measuring its
significance, we perform 1000 Monte Carlo simulations of our data, randomly shuffling the
velocities in the group and re-calculating the $\Delta$ statistic.  This allows us to calculate the
probability that our measured $\Delta$ can be obtained from a random distribution of recession
velocities.  

It is important to note that there are two insensitivities of the $\Delta$ statistic that must also
be considered. The first is related to the $\Delta$ statistic's inability to detect superimposed
substructures, mentioned by Dressler \& Shectman (1988).  A second caveat is that the $\Delta$
statistic is insensitive to equal mass mergers occurring in the plane of the sky (i.e. similar mean
velocity and dispersion), as discussed by Pinkney \etal (1996).

We find suggestions of substructure in the NGC\,5044 group, with the group's $\Delta$ value
significant at greater than the 99.9 percent level.  It is possible that high $\Delta$ values can be
obtained via rotation or velocity gradients across the group.  Based on this, we have examined our
velocity data for any signs of rotation, however we find no evidence for bulk rotation with
$\nu_{rot}\sim$53$\pm49$\kms.

\begin{figure*}
\centering
\includegraphics[scale=0.70,angle=-90]{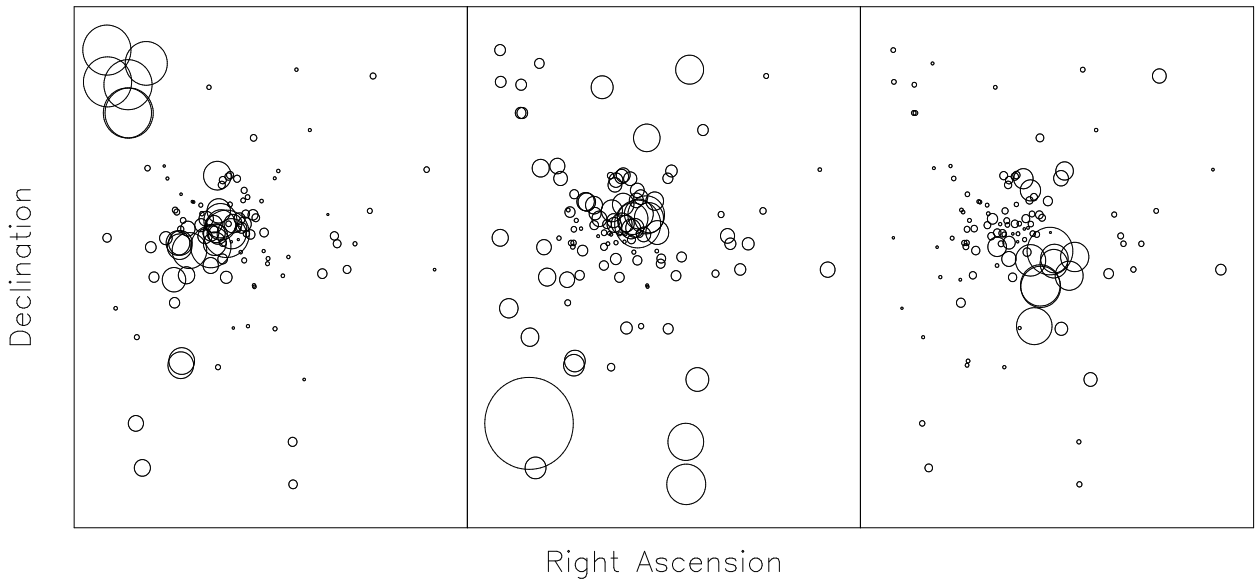}
\caption{Bubble diagram for the Dressler-Shectman substructure test.  Each circle represents a
galaxy in the NGC\,5044 group with sizes scaled by $\mathrm{e}^\delta$ (see text).  The left panel
shows galaxies scaled by their observed $\delta$ while the centre and right panels show the maximal
and median substructure results from our Monte Carlo simulations of the data.  Orientation and area
displayed in each panel are the same as in Fig. \ref{spatial}. }
\label{DS88}
\end{figure*}

To visually search for merging sub-groups that could be responsible for the large $\Delta$ value, in
Fig. \ref{DS88} we show a bubble plot for the NGC\,5044 group.  Here, each galaxy is represented
with a circle whose radius is scaled as $e^\delta$ (where $\delta$ is taken from Eq.  \ref{delta}).
As is evident from this figure, there is a significant sub-clump of galaxies in the north-east
outskirts of the group, which is responsible for the high $\Delta$ value.  Removing these galaxies
and re-running the substructure test results in the group's $\Delta$ value being significant at only
the 70 percent level.

\subsection{Number density profile}
\label{number density}

Hierarchical formation in the context of the $\Lambda$CDM paradigm makes particular statistical
predictions regarding the shape of resulting density profiles.  These predictions are primarily
motivated by numerical simulations and suggest that halo profiles are well described by a
combination of power laws growing progressively steeper with increasing radius (Dubinski \& Carlberg
1991; Navarro, Frenk \& White 1995, 1996, 1997).  Such exponential profiles have been found to be
applicable across a range of groups and clusters (e.g. Carlberg \etal 1997; Carlberg, Yee \&
Ellingson 1997; Biviano \& Girardi 2003).

The average projected number density of galaxies, $\Sigma_N$, can be represented as a projection of
the volume density $\nu (r)$ via an Abel integral where 

\begin{equation}
\label{vol_surf}
\Sigma_N(R) = 2\int_{R}^{\infty}\mathrm{\nu} (r)\frac{r}{\sqrt{r^2-R^2}}dr.
\end{equation} 

\begin{figure}
\centering
\includegraphics[scale=0.36,angle=-90]{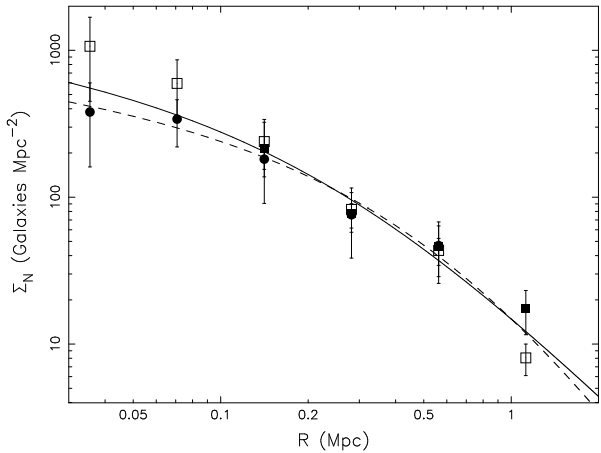}
\caption{Average projected number density as a function of radius.  Open squares, filled circles and
filled squares represent NED, AAOmega and 6dFGS data respectively.  Data are scaled to match in
number density at radii between 0.1 and 0.6\,Mpc (see $\S$\ref{number density}).  Note that as a
result of this scaling symbols overlap.  The solid and dashed lines represent the best fit NFW and
Hernquist profiles.} 
\label{surface_density}
\end{figure}

\noindent Here we adopt a general form of the volume density

\begin{equation}
\label{v_dens}
\nu (r) = \frac{A}{r(r+a_v)^p},
\end{equation}

\noindent where $a_v$ is the radius of the transition between the inner and outer profile slopes.
Different outer slopes of the profiles are obtained for a fixed valued of $p$, and adopting $p=2$ or
3 gives an NFW (Navarro, Frenk \& White 1995, 1996, 1997) or Hernquist (Hernquist 1990) profile
respectively.  In Fig. \ref{surface_density} we show the binned surface density profiles calculated
for each of the three data sets (either NED, 6dFGS or AAOmega), where the 6dFGS and AAOmega bins
have been corrected for their known incomplete spatial coverage.  As a rough correction to
facilitate fitting a mean density profile we have scaled the data to match in density between 0.2
and 0.6\,Mpc, applying the derived offset to the remaining bins.  

The profiles fit to this corrected data are shown with the solid and dashed lines (NFW and Hernquist
profiles) in Fig. \ref{surface_density}, where bins containing a single galaxy have been excluded
from the fitting process.  Both fits are reasonable, however the NFW profile provides a
statistically better fit.  The scale radius for the best fit NFW profile is $a_v=$0.26$\pm$0.02,
consistent with the CNOC cluster profile fits of Carlberg \etal (1997).  For the remainder of
relevant analyses we will consider only the best fit NFW profile. 

\subsection{Velocity dispersion profile}
\label{sigma_profile}

The underlying mass distribution of a group or cluster is related to the volume density and 
velocity dispersion profiles via the Jeans equations.  By measuring the observed density and velocity
distributions, one can infer properties of the group mass distribution.  This analysis is often
applied to group and cluster systems where galaxies can be treated as tracer particles, with the
caveat that interpretation using the Jeans equations is only appropriate for systems in equilibrium.

The velocity dispersion profiles of clusters are generally found to be falling with radius (e.g.
Carlberg, Yee \& Ellingson 1997; Carlberg \etal 1997; Girardi \etal 1998; Biviano \& Girardi 2003;
Biviano \& Katgert 2004) and indicative of centrally concentrated mass distributions.  Group studies
measuring velocity dispersion profiles are generally in good agreement, finding either falling or
roughly flat profiles with radius (e.g. Zabludoff \& Mulchaey 1998; B06).  However,
there is some evidence that the velocity dispersion is actually smoothly rising with radius
(Carlberg \etal 2001), suggesting that galaxies are concentrated with respect to the dark matter
distribution due to contraction within the halo (e.g. by dynamical friction). 

The velocity dispersion at a given radius is calculated here in a similar fashion to the number
density, using the projection of the radial velocity dispersion and number density such that

\begin{equation}
\label{psig}
\sigma_p^2(R) = \frac{1}{\Sigma_N(R)}\int_{R}^{\infty}\mathrm{\nu}\sigma_r^2
\left(1-\beta\frac{R^2}{r^2}\right)\frac{r}{\sqrt{r^2-R^2}}dr,
\end{equation}

\noindent where $\beta=1-\sigma_\theta^2/\sigma_r^2$ is the velocity anisotropy parameter.  We have
adopted a simple form of the radial velocity dispersion used by Carlberg \etal (1997) defined as 

\begin{equation}
\label{rsig}
\sigma_r^2(r) = \frac{B}{b+r}.
\end{equation}

\noindent where $B$ and $b$ are free parameters in our fit.

\begin{figure}
\centering
\includegraphics[scale=0.36,angle=-90]{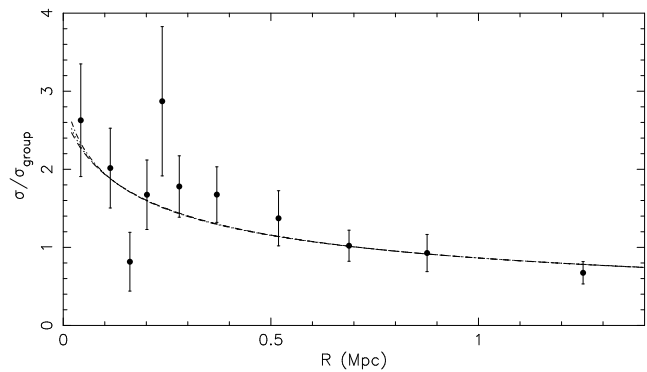}
\caption{Velocity dispersion profile of the NGC\,5044 group.  NGC\,5044 is assumed to be the centre
of the group.  Bins contain 10 galaxies each and 1$\sigma$ jacknife errors are shown.  Best fit
profiles for $\beta=1,\frac{1}{2}$ and 0 are overlayed (lines overlap).}
\label{v_proj}
\end{figure}

In Fig. \ref{v_proj} we show the velocity dispersion profile derived for our data, with best fit
velocity dispersion profiles for $\beta=1,\frac{1}{2}$ and 0 (purely radial, marginally isotropic,
and isotropic orbits).  Bins contain 10 galaxies each and velocity dispersions have been calculated
using the gapper algorithm (Beers \etal 1990; Eq.  \ref{gapper}).  As expected for a relaxed group,
we find the velocity dispersion falling with radius regardless of our allowed values of $\beta$,
implying that the group mass is centrally concentrated and in agreement with the profile found by B06
for their stacked GEMS groups.

There is some suggestion of fluctuation in the velocity dispersion profile between 100 and 200\,kpc
(also seen in Fig. \ref{vel_radius}), however due to our relatively low numbers its significance is
unclear.  Fluctuations in the velocity dispersion profile are commonly observed in clusters (see
e.g. den Hartog \& Katgert 1996) and studies which allow the orbital anisotropy parameter, $\beta$,
to vary find that these fluctuating profiles can be reasonably fit by a varying orbital distribution
with radius (see e.g. Biviano \& Katgert 2004; Hwang \& Lee 2007).  These velocity dispersion
variations can also be produced if there are multiple sub-populations in the group, either on
uniquely different orbits (e.g. radial or tangential) or offset from the assumed centre of the
group.

\subsection{Mass estimates}
\label{mass_estimates}

\begin{table}
\centering
\scriptsize
\caption{Comparison of mass estimates from the literature.}
\begin{tabular}{llrc}
\hline
Source				&Method		& $M$ (10$^{13}$M$_\odot$)	&$R$ (Mpc) \\	
(1)				&(2)		&(3)			&(4)	\\	
\hline                          	                                        
This Work			&Dynamical	& 9.2$\pm$1.6 		&\ldots	\\
Brough \etal (2006)		&Dynamical	& 6.9$\pm$1.9		&\ldots	\\
David \etal (1994)		&X-ray		& 1.6			& $<$0.18 \\
Betoya-Nonesa \etal (2006)	&X-ray		& 1.6			& $<$0.18 \\
Betoya-Nonesa \etal (2006)	&X-ray		& 5.1			& $<$0.58 \\
\hline
\end{tabular}
\label{mass} 
\end{table}	

Table \ref{mass} shows a comparison of mass estimates for the NGC\,5044 group in the literature,
derived both dynamically and from X-rays.  As this table shows, there are a variety of estimates for
the NGC\,5044 group mass from $\sim10^{13}$ to $10^{14}$\,M$_{\odot}$.  Given this variation, and our
attempt to compare the NGC\,5044 group with other groups and clusters, we will briefly examine the
reliability and consistency of these mass estimates. 

For this comparison we have built a sample of groups and clusters using two separate data sets.  The
first is the 15 Southern GEMS groups studied by B06, which contains groups in the range
$12.3<\mathrm{log}\,M_{\odot}<13.8$.  B06 have calculated the dynamical properties of these groups
(i.e. virial mass, radius etc.) and the X-ray properties of the groups have been presented by Osmond
\& Ponman (2004).

The second data set consists of 38 clusters from the {\it Advanced Satellite for Cosmology and
Astrophysics} (ASCA) Cluster Catalogue of Horner (2001; ACC).  The velocity dispersion and virial mass
measurements for these clusters have been adopted from Girardi \etal (1998), and only those clusters
with $>$30 confirmed members are included in this analysis.      

In order to compare the X-ray and dynamical masses for a number of sources, we first need normalise
them to a standard mass scale.  For dynamical masses this normalisation is not straight forward.
Masses for the GEMS and ACC samples are both calculated as virial masses, which implies a density
contrast of $\delta_{vir}=18\pi^2\approx178$ for an $\Omega_0=1$ universe (Bryan \& Norman 1998).
However there is no implicit radius assumption in these calculations, so scaling these masses to a
different density contrast (e.g. $\delta=200$) is difficult.

We calculate X-ray masses to a similar density contrast as the dynamical estimates.  This has been
done using the $\beta$-model fit parameters to calculate the group mass in terms of density contrast
and X-ray temperature (Horner, Mushotzky \& Scharf 1999) such that

\begin{eqnarray}
\label{xray_mass}
M(\delta,\beta,T_{\mathrm{X}})&=&1.1\times10^{15}\delta^{-1/2}\beta^{-3/2}\left(\frac{T_{\mathrm{X}}}
{\mathrm{keV}}\right)^{3/2} \nonumber\\ & &\times\left(1-0.01\frac{\delta r_c^2}{\beta
T_{\mathrm{X}}}\right)^{3/2}h^{-1}M_{\odot}.
\end{eqnarray}

\noindent In this case, we adopt $\delta=178$ and use the $\beta$-model fits from Osmond \& Ponman
(2004) for the GEMS groups and Fukazawa (1997) for the ACC cluster sample.

\begin{figure}
\centering
\includegraphics[scale=0.40,angle=-90]{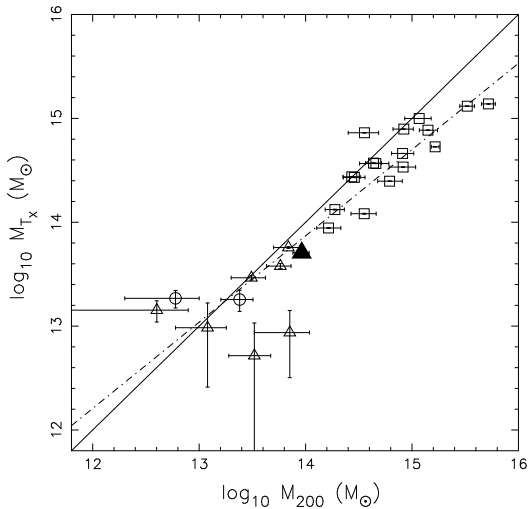}
\caption{Comparison between dynamically and X-ray derived mass estimates for groups and clusters.
Groups from B06 are shown with open triangles and circles, representing group and galaxy scale X-ray
emission respectively.  Open squares are Horner clusters with velocity dispersions measured by
Girardi \etal (1998; see text for details).  The solid line represents a one-to-one correlation, and
the dashed line is a best fit to the data with a slope of 0.83$\pm$0.08.  Our measurements for
NGC\,5044 are shown as the large, filled triangle.} 
\label{dyn_tx_mass}
\end{figure}

In Fig. \ref{dyn_tx_mass} we show the comparison of dynamical masses and re-calculated X-ray masses.
This comparison shows a systematic offset between X-ray and dynamical masses, however the two
estimates scale almost linearly.  For the NGC\,5044 group, the mass we calculate using Eq. \ref{xray_mass}
is consistent with that of Betoya-Nonesa \etal (2006), i.e. $\sim$5.1$\times$10$^{13}$\,M$_{\odot}$,
while our dynamical mass is consistent with the general group and cluster trend.  So, while the
dynamical and X-ray masses for this group disagree, their disagreement is consistent with the
systematic offset present between the two methods used.

\subsection{Group scaling relations}
\label{group_relations}

In self-similar models of group and cluster formation the amount of gas present in a halo is
directly related to the depth of the potential well, leading to correlations of X-ray luminosity
with both gas temperature and velocity dispersion.  In this self-similar model the X-ray luminosity
scales directly with velocity dispersion as $L_X\propto \sigma_{\nu}^4$.  At cluster masses there is
general agreement that observations are consistent with the self-similar picture, with numerous
studies finding power law slopes consistent with $\sim$4 (e.g. Mulchaey \& Zabludoff 1998; Helsdon \&
Ponman 2000a; Mahdavi \& Geller 2001; Ortiz-Gil \etal 2004; Hilton \etal 2005).  At group masses,
however, there is some disagreement as to whether or not the $L_X$--$\sigma_{\nu}$ relation is
merely extension of the $\sigma_{\nu}^4$ relation found for clusters, or shallows to a slope of
$\sim$2.4 (e.g. Helsdon \& Ponman 2000b; Xue \& Wu 2000).

Osmond \& Ponman suggest that, rather than an actual flattening in the $L_X$--$\sigma_{\nu}$
relation, the observed break at group masses could be a result of $\sigma_{\nu}$ poorly tracing the
depth of the potential well.  Conversely, Plionis \& Tovmassian (2004) argue that the observed
flattening in the $L_X$--$\sigma_{\nu}$ relation at low velocity dispersions is spurious, and is
only observed as a result of a statistical bias in sampling of the low-$L_X$ limit.

Theoretically, one can predict not only the $L_X$--$\sigma_{\nu}$ correlation, but also the
relationship between total mass of a system and its X-ray temperature or luminosity.  Following
self-similar arguments one can show that $M\propto T_X^{3/2}$ (Afshordi \& Cen 2002) and using
$L_X\propto T_X^2$, arrive at $L_X\propto M^{3/2}$.  As with the $L_X$--$\sigma_{\nu}$ relation,
there is disagreement as to the consistency of observational data with self-similar predictions.
Horner, Mushotzky \& Scharf (1999) find a slope for their cluster sample consistent with the
predicted power law slope of $\sim$1.5 in the $M$--$T_X$ relation, while Nevalainen \etal 2000 and
Finoguenov, Reiprich \& B\"ohringer (2001) find suggestions of a steeper relation, owing to the
inclusion of groups in their X-ray samples.  This slope change can be attributed to either an
increase in the intrinsic scatter of the $M$--$T_X$ relation at group masses or a break in the
relation where group and clusters follow fundamentally different trends.  

While it is outside the scope of this paper to examine the nature of the $L_X$--$\sigma_{\nu}$ and
$L_X$--$M$ relations in detail, it is of interest to place the NGC\,5044 group in this parameter
space relative to other groups and clusters.  Here we adopt the GEMS and ACC samples described in
$\S$\ref{mass_estimates}.  

\begin{figure}
\centering
\includegraphics[scale=0.34,angle=-90]{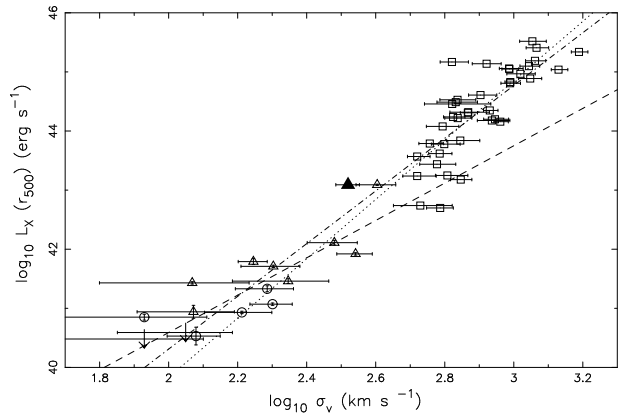}
\caption{X-ray luminosity variation ($L_X (r_{500})$) with velocity dispersion ($\sigma_{\nu}$) in
groups and clusters.  Group and galaxy scale emission from B06 groups are shown using triangles and
circles respectively.  X-ray undetected groups are shown as upper limits.  Open squares are Horner
clusters with velocity dispersions measured by Girardi \etal (1998; see text for details).  Our
measurement of the NGC\,5044 is plotted as the large, filled triangle.  Dotted, dashed and
dot-dashed lines represent best fits to the cluster, group and combined samples respectively.} 
\label{lx_sig}
\end{figure}

\begin{figure}
\centering
\includegraphics[scale=0.34,angle=-90]{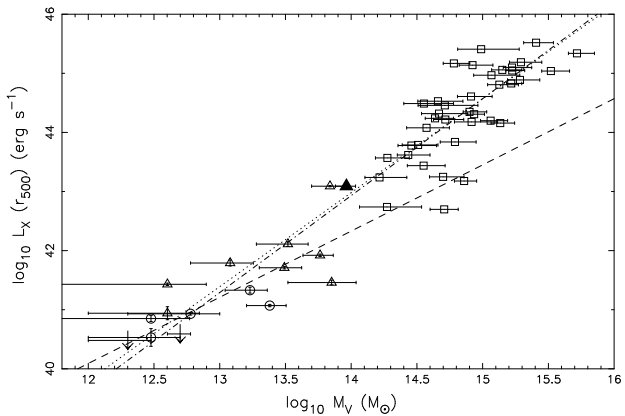}
\caption{X-ray luminosity variation with dynamical mass in the groups and clusters. Symbols are the
same as Fig. \ref{lx_sig}.  Lines are best fits to the $L_X-M$ data and dotted, dashed and
dot-dashed lines represent fits to cluster, group and combined data respectively.  Dynamical masses
from Girardi \etal (1998) have been adjusted to $\delta$=178 (see $\S$\ref{mass_estimates}).} 
\label{lx_mv}
\end{figure}

In Figs. \ref{lx_sig} and \ref{lx_mv} we show the X-ray luminosity--velocity dispersion and
--dynamical mass relations for the combined GEMS and ACC sample, with best fits to the GEMS, ACC and
combined samples overlayed.  The NGC\,5044 group is fit well by both the cluster and combined sample
fits in the $L_X$--$\sigma_{\nu}$ relation.  While our fits suggest a difference in the group and
cluster relations, the velocity dispersion measurements for low mass groups are highly uncertain
(owing to their low number of member galaxies).  The fits we find for the $L_X$--$M_{200}$ relation
to the individual GEMS and ACC samples are consistent with the idea of a shallowing slope at group
masses. As in the $L_X$--$\sigma_{\nu}$ relation, the NGC\,5044 group is better fit by the ACC or
combined sample trends.

Our results therefore suggest that, while the NGC\,5044 group's mass of $\sim10^{14}$\,M$_{\odot}$
places it between the group and cluster regimes, the relationship of its X-ray and dynamical
properties show it to be more akin to clusters than groups.

\section{Galaxy properties}
\label{galaxy}

\subsection{Galaxy colours} 
\label{colours} 

With numerous studies of the photometric properties of early-type galaxies, it has been
established that brighter, more massive galaxies are increasingly redder in colour (e.g. Visvanathan
\& Sandage 1977; Baldry \etal 2004; Blanton \etal 2005).  Further comparison of colour-magnitude
relations (CMRs) between different clusters have shown nearly identical results (see, e.g. Bower,
Lucy \& Ellis 1992).  In fact, Hogg \etal (2004) used a sample of SDSS galaxies to show that the
slope of the CMR is roughly constant across a wide range of environments, a result that has been
confirmed using larger samples from the SDSS (e.g.  Gallazzi \etal 2006).

More detailed studies on the physical origin of the CMR have focused on comparing the stellar
populations of galaxies at differing magnitudes (e.g. Bressan, Chiosi \& Tantalo 1996; Gallazzi \etal
2006) and have found that the CMR is primarily a mass--metallicity relation.  Observed trends of
increased metallicity towards high masses are consistent with feedback mechanisms, such as supernova
driven winds that will be increasingly efficient at blowing gas out of low mass halos and
suppressing star formation from enriched gas (e.g. Trager \etal 2000; Thomas \etal 2005).
The increased blue-ward scatter about the CMR at low masses is dominated by age effects (Gallazzi
\etal 2006; Kodama \& Arimoto 1997), where lower mass galaxies are still observed to be forming
stars and higher mass galaxies are predominantly passively evolving (i.e. the ``downsizing''
effect).

\begin{figure}
\centering
\includegraphics[scale=0.34,angle=-90]{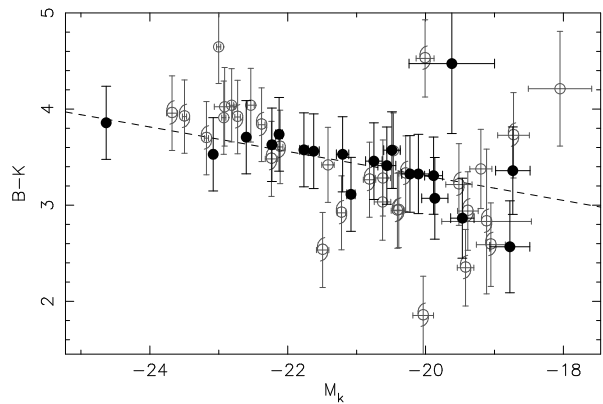}
\caption{B-K vs. M$_K$ relation for galaxies in the NGC\,5044 group.  Elliptical, S0 and late-type
galaxies are represented by filled circles, open circles and spirals respectively.  K-band errors
are taken from either the 2MASS and P00 catalogues, for B-band error estimates we adopt the
scatter around the best fit in Fig. \ref{b_phot_comp} of 0.37\,mags.  The solid line is the best fit
CMR to the early-type data in the sample. }
\label{cmr}
\end{figure}

In Fig. \ref{cmr} we show the $B-K$ vs. M$_K$ colour-magnitude relation derived for our NGC\,5044
sample.  Of the four galaxies with very red colours ($B-K>4.6$), the two S0's and spiral galaxy
(open circles and spirals in Fig. \ref{cmr} respectively) are seen edge on, so we expect increased
reddening from dust in the disk.  The fourth (elliptical) galaxy has a very large error associated
with its K-band magnitude, and we suspect that the galaxy's very red colour is due to this
uncertainty.  

A linear fit to the early-type galaxies gives a slope of $-0.13\pm0.03$ and intercept of
$0.76\pm0.55$ (dashed line in Fig. \ref{cmr}), consistent with previous analyses of $B-K$ CMRs (e.g.
Mobasher, Ellis \& Sharples 1986; Forbes \etal 2008).  This also shows that dwarf galaxies in our
sample are, on average, $\sim$0.4 mags bluer that the most massive galaxies.

\begin{figure}
\centering
\includegraphics[scale=0.34,angle=-90]{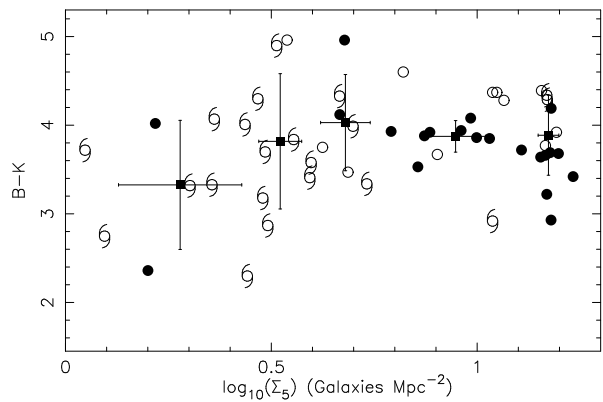}
\caption{B-K vs. local density ($\Sigma_5$) for galaxies in the NGC\,5044 group. Symbols are the
same as in Fig. \ref{cmr}.  Squares show the mean density and colour for every 10 galaxies with
1$\sigma$ errors.}
\label{local_density}
\end{figure}

The correlation of galaxy colours with their local density, where redder galaxies are found
predominantly in higher density environments, is now a well established trend (Dressler 1980;
Kauffmann \etal 2004; Smith \etal 2005; Blanton \& Berlind 2007).  Recent studies using large 2dFGRS
and SDSS samples have found increasing evidence that local density not only correlates with galaxy
colour, but also galaxy star formation properties (e.g.  Lewis \etal 2002; G\'omez \etal 2003).  For
our sample, we calculate local density for each galaxy, $\Sigma_5$, using the projected distance of
the 5$^{th}$ nearest neighbour galaxy.  The relationship of $B-K$ colour with $\Sigma_5$ is shown in
Fig. \ref{local_density}.  We note that low density outskirts of the NGC\,5044 group are the most
poorly sampled, and so density measurements in these regions are likely an underestimate of the true
density.  Nevertheless, this figure certainly suggests agreement with the findings of Lewis \etal
(2002) and Gomez \etal (2003), that above a certain local density (in this case
$\Sigma_5=2-3$\,galaxies Mpc$^{-2}$) galaxies transition from blue and star-forming, to passively
evolving and red.

\subsection{Spatial distribution of galaxies and segregation}
\label{spatial_dist}

With our large sample of confirmed 111 group members it is possible for us to explore the spatial
and kinematic distribution of galaxies in significantly more detail than previous group studies.
The most well known examples of spatial segregation, the morphology-density and morphology-radius
relations (e.g. Dressler 1980; Butcher \& Oemler 1984) have been well studied in galaxy clusters
owing to the large numbers of galaxies typically available relative to the group environment.
However studies focusing on galaxy groups also find significant evidence for segregation.  In
addition to spatial segregation, galaxies in clusters are found to be dynamically segregated with
respect to both luminosity and morphology in simulations (Yepes \etal 1991; Fusco-Femiano \& Menci
1998), which is supported by observations of dynamical segregation in both groups and clusters
(Girardi \etal 2003; Lares \etal 2004). 

Here we examine evidence for both spatial and dynamical segregation amongst our galaxy group sample.

\subsubsection{Spatial segregation}

\begin{figure}
\centering
\includegraphics[scale=0.40,angle=-90]{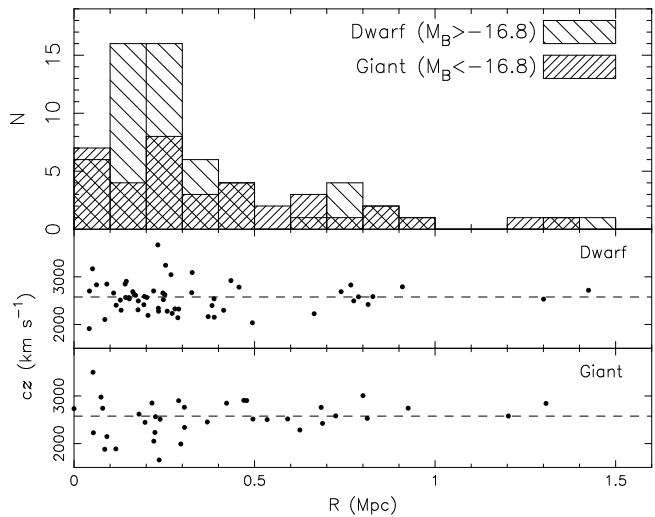}
\caption{Histogram showing number of galaxy type, either dwarf or giant, with projected radius.
Second and third panels show recession velocity vs. projected radius for dwarf and giant sub-samples
respectively.  Dashed lines in panels two and three show the mean group velocity of 2577\kms.}
\label{dg}
\end{figure}

\begin{figure}
\centering
\includegraphics[scale=0.40,angle=-90]{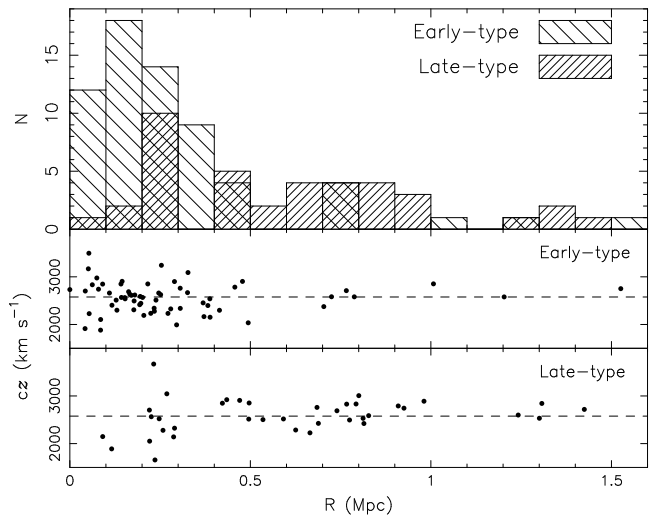}
\caption{Same as Fig. \ref{dg}, except galaxies are divided by morphology (early- vs. late-type)
rather than luminosity.} 
\label{el}
\end{figure}

To address the question of spatial segregation we subdivide our total sample by morphology (early-
vs. late-types) and luminosity (where $M_B=-16.8$\,mag is used as the break between dwarf and giant
galaxies), the results of which are shown in Figs. \ref{dg} and \ref{el}.

When split by luminosity (Fig. \ref{dg}), galaxies show no significant evidence for a radial bias,
i.e. the two luminosity sub-samples are similarly distributed with radius.  The hypothesis that
dwarf and giant galaxies have the same radial distribution is only marginally rejected at the 70
percent level by a Kolmogorov-Smirnov (KS) test, confirming the visually apparent lack of a strong radial bias with
luminosity.  This is consistent with the known trend in dwarf-to-giant ratios (DGR) for groups and
clusters, where the DGR is independent of position observed in the group (see e.g.  FS91).

Splitting galaxies based on morphology (Fig. \ref{el}) gives a markedly different result in that
galaxies are differently distributed at the 5 $\sigma$ level as determined using a KS test.  This is
consistent with expectations from the morphology-density and morphology-radius relations (Butcher \&
Oemler 1976; Dressler 1980) and our findings from $\S$\ref{colours}. 

It is important to keep in mind that our group sample is assembled from a variety of sources with
differing spatial coverage, selection and completeness (see Fig. \ref{spatial} and $\S$\ref{data}) ,
all of which could be inducing a radial bias in the results presented in Figs. \ref{dg} and
\ref{el}.  In particular, the region within $\sim$600\,kpc of the group center (approximately the
area covered by FS90 and, therefore, our AAOmega observations) includes significant numbers of faint
(dwarf) galaxies that are likely not included in the samples at larger radii.  Repeating the KS test
on only galaxies within the central 600\,kpc gives similar results to those above; dwarf and giant
galaxies appear similarly distributed (at the $\sim$60 percent level) while early- and late-type
galaxies show a difference in their radial distributions significant at the 3 $\sigma$ level.

\subsubsection{Dynamical segregation}

The dynamical segregation of galaxies in the NGC\,5044 group has been examined previously by Cellone
\& Buzzoni (2005) using a sample of 26 early- and late-type galaxies.  When dividing their sample
based on morphology, they find early- and late-type galaxies exhibit velocity dispersions that
differ at the 93 percent level (as measured using an $F$ test).  Cellone \& Buzzoni (2005) find no
evidence for luminosity segregation in their early-type galaxies, however were unable to examine
luminosity segregation between their early- and late-type samples due to low numbers of late-type
galaxies.  

Here we revisit the question of dynamical segregation in the NGC\,5044 group using the morphology
and luminosity sub-samples described above.  We then compare the line-of-sight velocity
distributions of sub-samples using two different methods to look for dynamically distinct behaviour.
The first comparison is carried out using a KS test, which computes the likelihood of the two
different samples being drawn from the same parent velocity distribution.  The second test we use is
an F-test, which gives the likelihood of the two distributions having significantly
different variances.  Table \ref{disp_tab} summarises the results of the KS and F-test for our
luminosity and morphology sub-samples. 
 
Our results for morphological segregation computed using the F-test are consistent with Cellone \&
Buzzoni (2005), however increasing the sample of galaxies from 26 to 105 has decreased the
significance slightly.  There is marginal evidence of a variation in the distributions of early-type
dwarf vs. giant galaxies and amongst early- vs. late-type dwarf galaxies, however these results are
of low statistical significance ($\sim 1.6\sigma$).

When split by either luminosity or morphology our sample of NGC\,5044 group galaxies is remarkably
homogeneous, showing little strong evidence for any dynamical segregation.  Results from the KS test
show no evidence that any of the sub-samples are statistically different distributions.  In fact,
the results for early- vs. late-type galaxies seem to suggest that the two distributions are
actually {\it similar} at the $\sim2\sigma$ level. Giant ($M_B<-17$) galaxies in our sample are
similar with respect to their dynamics in the group irrespective of morphological type. 

\begin{table*}
\footnotesize
\caption{Results from tests for dynamical segregation.}
\begin{tabular}{lcrrccccrr}
\hline
Sample &  Sub-sample  &	N$_1$	&N$_2$	& $\bar{\mathrm{c}z_{1}}$ &  $\sigma_{1}$ & $\bar{\mathrm{c}z_{2}}$ &  $\sigma_{2}$ & $L_{\mathrm{KS}}$	& $L_{\mathrm{F}}$ \\
 & & &	& (\kms)  & (\kms)  & (\kms) & (\kms)  & 	&  \\
 (1)  &(2) &(3)    & (4)  & (5)  & (6) & (7)  &(8)         & (9)& (10) \\
\hline
\multicolumn{10}{|c|}{{\it Luminosity Segregation}}\\
M$_B<-16.8$ vs. M$_B>-16.8$ 	& all morphologies	& 37 & 58 & 2542$\pm$71 & 397$\pm$71 & 2540$\pm$40 & 302$\pm$36 & 46.1 & 84.8 \\
M$_B<-16.8$ vs. M$_B>-16.8$ 	& late-type    		& 18 & 17 & 2492$\pm$108& 438$\pm$104& 2580$\pm$77 & 317$\pm$64 & 27.6 & 8.7 \\
M$_B<-16.8$ vs. M$_B>-16.8$ 	& early-type   		& 19 & 41 & 2582$\pm$79 & 380$\pm$67 & 2522$\pm$53 & 296$\pm$38 & 67.8 & 91.9 \\
\multicolumn{10}{|c|}{{\it Morphological Segregation}}\\
early- vs. late-types 		& all luminosities 	& 60 & 35 & 2542$\pm$48 & 317$\pm$38 & 2547$\pm$61 & 380$\pm$59 & 16.7 & 90.8 \\
early- vs. late-types 		& M$_B>-16.8$        	& 41 & 17 & 2522$\pm$53 & 296$\pm$38 & 2580$\pm$77 & 317$\pm$64 & 76.4 & 94.0 \\
early- vs. late-types 		& M$_B<-16.8$        	& 19 & 18 & 2582$\pm$79 & 368$\pm$67 & 2492$\pm$108& 438$\pm$104& 41.6 & 8.5 \\
\hline
\end{tabular}
\flushleft Notes.  Columns are as follows: (1) and (2) describe the samples being compared. (3) and
(4) give the number of galaxies in each sub-sample.  (6)--(8) give the mean velocities and
dispersions of the two different sub-samples.  (9) is the significance, from a K-S Test, that the
two distributions are drawn from different distributions.  (10) gives the significance, via an
F-test, that the distributions have different variances. 
\label{disp_tab}
\end{table*}

\section{Discussion and conclusions}
\label{discussion}

We have used multi-object spectroscopy to obtain recession velocity measurements for galaxies in the
NGC\,5044 group.  Combining these new observations with available data from literature we are able
to define the NGC\,5044 group as containing 111 members with $M_B\leq-13.5$\,mags, nearly a
three-fold increase over previous numbers of confirmed group members.

An analysis of common dynamical indicators such as crossing time, line-of-sight velocity
distribution, position of the X-ray peak relative to the centroid suggest that the NGC\,5044 group
is relaxed and virialized despite the observed 150\kms peculiar velocity of the brightest group
galaxy NGC\,5044.  This conclusion of virialization is consistent with XMM X-ray contours for the
group, which are very regular and undisturbed.  We note, however, that none of our tests for
virialization are sensitive to effects along the line-of-sight, and so the true dynamical state of
the group remains somewhat unknown.  Taking the above indicators at face value, however, the group's
virialization suggests that the NGC\,5044 group has experienced no {\it major} sub-group mergers in
several crossing times ($\sim$1\,Gyr). 

While the dynamical indicators discussed above will give hints as to the timescale of major merger
activity, it is likely that low-mass sub-group mergers will not significantly disrupt the
virialization of the system as determined via these indicators.  By computing the Dressler-Shectman
$\Delta$ statistic, we have been able to visually and statistically search for space-velocity
substructure in the group.  In doing so, we find evidence for a low mass substructure $\sim$1.4\,Mpc
from NGC\,5044 group centre.  

Two body interactions are expected to lead to dynamical and luminosity segregation in mature
groups such as NGC\,5044 (e.g.  Fusco-Femiano \& Menci 1998; Lares \etal 2004).  In examining the
NGC\,5044 group's galaxy population however, we find that galaxies are primarily segregated with
respect to their morphologies, and there is no {\it strong} evidence for segregation in either
dynamics or luminosity.

While luminosity segregation is an expected outcome of mergers in groups (e.g. Fusco-Femiano \&
Menci 1998; Lares \etal 2004), Ludlow \etal (2007) have used N-body simulations to show that it is
possible to eject sub-halos from a group, via multi-body interactions, resulting in significant
numbers of associated sub-halos\footnote{Ludlow \etal (2007) define associated halos as those that
have, at some point, passed within the virial radius of the central halo} residing in the outskirts
of the group (as far as 5 times the virial radius).  In relaxed groups such as NGC\,5044, galaxies
have had significant time for two-body interactions to take place, and so the ejection of galaxies
from the group centre is more likely to wash out segregation trends.  In addition, this complicates
the common interpretation of group-centric radius as a tracer of accretion history as it is no
longer clear that galaxies in the outskirts are actually the most recent galaxies to have been
accreted into the group potential.

Early-type galaxies in the NGC\,5044 group are well described by a linear $B-K$ colour-magnitude
relation, and are consistent with previous interpretations for the slope in the CMR due to
decreasing metallicity or age at lower galaxy masses.  Correlations of $B-K$ colour with local
density seem to further suggest that the majority of these bluer, fainter galaxies are residing in
the outskirts of the group, where densities are less than $2-3$\,galaxies Mpc$^{-2}$.  This is
consistent with findings for groups and cluster outskirts in the 2dFGRS and SDSS surveys (e.g.
Lewis \etal 2002; Gomez \etal 2003), however our low sample size prohibits a more quantitative
analysis of this effect.       

Future work will include a consideration of stellar populations in the galaxies of the NGC\,5044
group, measured using a combination of our AAOmega and 6dFGS spectra.  Combining the dynamical and
kinematic data of the confirmed group members presented here with age and metallicity measurements
will allow us to approach the analysis of this group and its galaxy population in significantly more
detail than previously attainable.  In particular, the expectation of pre-processing in groups
implies a necessary chemodynamical distinction between galaxies associated with the group and field
galaxies entering the group environment for the first time.

\section*{Acknowledgements}

The authors would like to thank Sergio Cellone for his helpful comments and suggestions.
We acknowledge the analysis facilities provided by {\sc iraf}, which is distributed by the National
Optical Astronomy Observatories, which is operated by the Association of Universities for Research
in Astronomy, Inc., under cooperative agreement with the National Science Foundation.  

This publication makes use of data products from the Two Micron All Sky Survey, which is a joint
project of the University of Massachusetts and the Infrared Processing and Analysis
Center/California Institute of Technology, funded by the National Aeronautics and Space
Administration and the National Science Foundation.  We also thank the Australian Research Council
for funding that supported this work.

\begin{appendix}
\section{Dynamical Formulae}
\label{Dynamical Formulae}

Virial mass, $M_{\mathrm{V}}$, has been calculated using the virial mass estimator from Heisler \etal (1985)

\begin{equation}
\label{m_vir_eqn}
M_{\mathrm{V}}=\frac{3 \pi N}{2 G} \frac{\sum_i \nu_{i,gc}^2}{\sum_{i<j}1/R_{gc,ij}},
\end{equation}

\noindent where $\nu_{i,gc}$ is the velocity of galaxy $i$ with respect to the group centroid and $R_{gc,ij}$ is the
projected group-centric separation from other galaxies. 

Velocity dispersion is calculated using the biweight scale estimator (Beers \etal 1990)

\begin{equation}
\label{sig_eqn}
\sigma_{\nu}=n^{1/2}\frac{\left[\sum_{\left|u_i\right|<1}(\nu_i-\bar{\nu})^2(1-u_i^2)^4\right]^{1/2}}
{\left|\sum_{\left|u_i\right|<1}(1-u_i^2)(1-5u_i^2)\right|},
\end{equation}

\noindent where $\nu_i$ denotes the recession velocity of the $i^{th}$ galaxy and $\bar{\nu}$ is the median group
velocity $u_i$ is defined as

\begin{equation}
\label{u_i_eqn}
u_i=\frac{(\nu_i-\bar{\nu})}{c\,\mathrm{MAD}}, 
\end{equation}

\noindent with MAD being the median absolute deviation (median deviation of the sample with respect to the
sample median).  Where we calculate the velocity dispersion for low numbers of galaxies ($<15$), we
use the gapper algorithm of Beers \etal (1990),

\begin{equation}
\label{gapper}
\sigma_\nu = \frac{\sqrt{\pi}}{n(n-1)}\sum_{i=1}^{n-1}w_ig_i,
\end{equation}

where $w_i=i(n-i)$ and $g_i=z_{i+1}-z_i$.  This has been shown to be more efficient for low sample
sizes than the biweight estimator (Beers \etal 1990).

The r$_{500}$ radius has been calculated using the velocity dispersion via

\begin{equation}
\label{r500}
r_{500}=\frac{0.096\sigma_{\nu}}{H_0}.
\end{equation}

The crossing time is defined as

\begin{equation}
\label{tc_eqn}
t_c=\frac{3 r_H}{5^{3/2}\sigma_{\nu}},
\end{equation}

\noindent in units of H$_0^{-1}$.  $r_H$ is the mean harmonic radius (Ramella \etal 1989) given by

\begin{equation}
\label{rh_eqn}
r_H=\pi\,D\,\mathrm{sin}\left[\frac{1}{2}\frac{n(n-1)}{2\sum_i\sum_{j>i}\theta_{ij}^{-1}}\right],
\end{equation} 

where $D$ represents the distance to the group and $\theta_{ij}$ is the pojected angular
separation between galaxies $i$ and $j$. 

\end{appendix}

\end{document}